\documentclass[
 reprint, amsmath,amssymb, aps,
]{revtex4-1}

\usepackage{graphicx}
\usepackage{dcolumn}
\usepackage{bm}

\newcommand{\be}{\begin{equation}} \newcommand{\ee}{\end{equation}}
\newcommand{\bea}{\begin{eqnarray}} \newcommand{\eea}{\end{eqnarray}}

\begin{document}

\title{Pattern dynamics of interacting contagions}

\author{Li Chen}
\affiliation{School of Physics and Information Technology, Shaanxi Normal University, Xi'an 710062, China}
\affiliation{Beijing Computational Science Research Center, 100193 Beijing, China}
\affiliation{Robert Koch-Institute, Nordufer 20, 13353 Berlin, Germany}
\email{chenl@snnu.edu.cn}

\date{\today}

\begin{abstract}
The spread of infectious diseases, rumors, fashions, innovations are complex contagion processes, embedded both in networked and spatial contexts. Here we investigate the pattern dynamics of a complex contagion, where two agents, say $A$ and $B$, interact with each other and diffuse simultaneously in the geographic space. The contagion process for each follows the classical susceptible-infected-susceptible kinetics, and their interaction introduces a potential change in the secondary infection propensity compared to the baseline reproduction ratio $R_0$. We show that nontrivial spatial infection patterns arise, when the susceptible move faster than the infected and the interaction between the two agents is neither too competitive nor too cooperative. Interestingly, the system exhibits pattern hysteresis phenomena that quite different parameter regions allowing for patterns exist in the direction of increasing $R_0$ and in the direction of eradication by its reduction. The latter shows a remarkable enhancement in the contagion prevalence, meaning that the infection eradication now becomes extremely difficult compared to the single-agent scenario and to the coinfection without space. Linearization analysis supports our observations, and we identified the required elements and dynamical mechanism behind the emergence of a pattern. These findings call for further investigation for their close relevance, both in biological and social contagions.
\end{abstract}

\pacs{05.45.Xt, 89.75.Hc, 87.23.Cc}
\maketitle

\section{Introduction}
After entering the new millennium, infectious diseases appear to be more active than ever, along with many new emerging pathogen strains. Well-known examples include SARS (Severe Acute Respiratory Syndrome) in 2003~\cite{hufnagel2004,colizza2007}, influenza A (H1N1) in 2009~\cite{fraser2009}, MERS (Middle East Respiratory Syndrome) coronavirus in 2012~\cite{zumla2015}, Ebola in 2013 \cite{ebola2015}, and the continuing H7N9 of avian influenza virus \cite{wang2017} \emph{etc}. To understand their contagion processes, mathematical models are an essential tool and have a long tradition in scientific communities that can date back to Bernuolli's work on smallpox vaccination in 1760 \cite{bernoulli1760}. Up to now, modeling effort amounts to be fruitful at all levels \cite{keeling2008,pastor2015}, ranging from very conceptual models \cite{kermack1927,hethcote2000} that capture the generic features of contagions, network models \cite{pastor2001,brockmann2006,balcan2009,belik2011} that focus on the underlying structure of population or commuting patterns, to very sophisticated computational models \cite{eubank2004,van2011}, where a variety of high resolution data like demographics, transportation, epidemiological features, and behavioral response \cite{gross2006} are incorporated. 

One important research line aims to understand more complicated contagions, where e.g. more than one pathogen is considered that circulate simultaneously in the population. This sort of complex contagion is motivated by the fact that the spread of different agents in the real world are not entirely independent, they often interact with each other \cite{sanz2014}. Well-known examples include the case of pneumonia bacterium like \emph{Streptococcus pneumoniae} and viral respiratory illness (e.g. seasonal influenza) where they mutually facilitate each other's contagion \cite{smith2013,shrestha2013}, and the coinfection between HIV (human immunodeficiency virus) and a host of other infections \cite{alter2006,abu2006,singer2009,pawlowski2012,chang2013}. The interaction among different agents can be either \emph{competing} \cite{newman2005,karrer2011,funk2010,marceau2011,poletto2013,sahneh2014,poletto2015} -- they suppress each other's circulation, or be \emph{cooperative} \cite{chen2013,cai2015,hebert2015,grassberger2016,janssen2016,chen2016,azimi2016,cui2017} -- they support each other's infection. The mean field treatment and simulations of structured population within the framework of percolation have revealed a rich spectrum of new dynamical features that are unexpected in the classic scenario of single contagion. For example, when different agents are in competing scenario, both one-pathogen-dominance and coexistence are possible, depending on the properties of involving pathogens and the underlining networks \cite{newman2005}. By contrast, in cooperative contagions avalanche (discontinuous) outbreak transition presents \cite{chen2013,cai2015}, along with many interesting spreading features such as its favor in clustered networks \cite{hebert2015}, first-order phase transitions in contagion prevalence \cite{chen2016} \emph{etc}.

\begin{figure*}
\includegraphics[width=1.5\columnwidth]{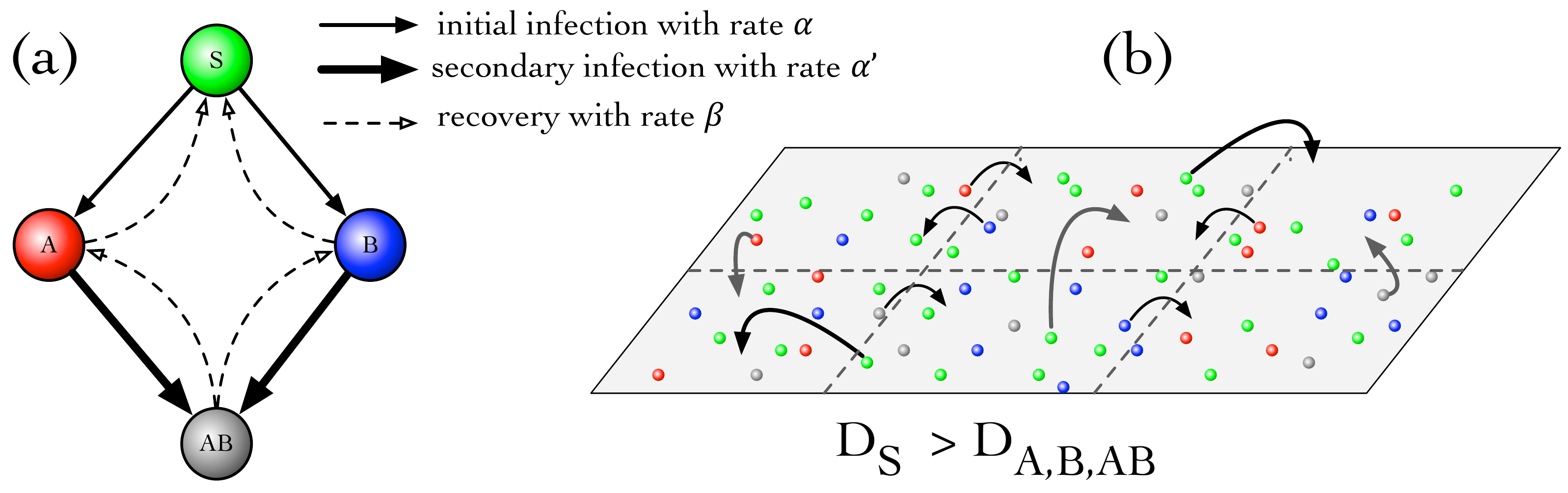}
\caption{\label{fig:1}\textbf{Model of interacting contagions.} (a) \emph{Mean field model}: consider two agents A and B, circulate in a given population, four states are then possible for host individuals -- the susceptible $S$, the partially infected $A$ or $B$, and the coinfected state $AB$. In the contagion process, $S$ becomes partial infected $A/B$ with the initial infection rate $\alpha$ by contacting the infected; the partially infected can be further infected by the other agent to be doubly infected $AB$ with the secondary infectious rate $\alpha'$. All infected individuals can recover by removing the agents with rate $\beta$. (b) \emph{Spatially coupled subpopulations}: when subpopulations are coupled through their geometrical neighborhood, the diffusion captures the local mobility of individuals, thus also the infectious agents they carry. Generally, the mobility of a given individual within the population depends on its dynamical state, e.g. in the epidemic spreading, the susceptible move faster than those infected, who might prefer to stay at home or in hospital for recovery. Mathematically, this means the diffusion coefficients satisfy $D_S>D_{A,B,AB}$ in reaction-diffusion framework described by Eq. (\ref{eq:rd}).}
\end{figure*}

Although these studies provide new insights into the temporal dynamics of complex contagion involving agent-agent interaction, their spatial behavior is largely unknown without explicitly incorporating the geographical dimension \cite{ostfeld2005}. The investigation of the spatial role in any case is indispensable for a full comprehension of contagion complexity \cite{murray2003,wang2014}, not only for its conceptual significance but also for its practical relevance in the real world \cite{brockmann2013}. Abundant empirical evidence has revealed nontrivial dynamics in spatial epidemiology, such as traveling waves \cite{grenfell2001}, infection patterns \cite{grassly2006}, and even spatial-temporal chaos \cite{gupta1998}. A major modeling effort in this regard is devoted to the study of traveling wave for understanding the infection propagation in geographic space, like the Black Death in Europe or the rabies epizootic in France \cite{murray2003}. The emergence of infection pattern received much less attention, yet a few mechanisms are proposed for pattern generation \cite{sun2016,cross1993}. These studies mainly focus on simple infection with a single agent, but incorporating additional compartments and/or additional dynamical processes. Till now, to our knowledge there is very rare work discussing the spatial dynamics of complex contagions, especially the possibility of pattern emergence. Only in Ref. \cite{chen2016}, a preliminary investigation was made to study the spatial dynamics of two interacting contagions assuming all individuals being of identical mobility (\emph{i.e.} all with the same diffusion coefficient), where novel propagation wave modes are revealed like receding fronts and standing waves. However, in realistic cases, individuals in different states are generally of different mobilities, the mobility of a given individual depends on her/his dynamical state. For example, the healthy people normally move faster than those sick who might prefer staying at home or in hospital for recovery. So \emph{what's the generic spatial dynamics when more than one agent diffuses in the population?} especially in the case when individuals in different dynamical state move differently. This question is also of particular interest in ecology community, where different diffusivities of species are thought to be responsible for the emergence of patchiness \cite{mimura1978}. In addition, recent works shows that multiplex networks as the underlying medium provide another mechanism for  generating patterns even if all species are of the same mobility \cite{nakao2010,gomez2013,kouvaris2015,nicolaides2016}.

In this work, we study the dynamical properties of two interacting SIS (Susceptible-Infected-Susceptible) agents in spatially extended context within the reaction-diffusion framework, see Fig.~\ref{fig:1}. When the susceptible agents are assumed to diffuse faster than the infected, we found that infection patterns emerge in a wide range of parameters. Counterintuitively, neither competition nor cooperation between the two agents is required to trigger the pattern formation, implying a rather loose precondition for the emergence. As we shall see, the linearization analysis of the system provides a good prediction, where positive eigenvalues imply instability modes, corresponding to pattern formation.

The paper is organized as follows: In Sec. \ref{Sec:MD}, we first briefly review the mean field treatment of SIS coinfection in Ref. \cite{chen2016}, which is the starting point of this study; Then the spatial model of interacting contagions is precisely defined in the framework of reaction-diffusion system. Main results are shown in Sec. \ref{Sec:PF}, where the impact of contagion interaction, the baseline reproduction ratio, the mobility of different compartments, are studied. Special interest goes to pattern hysteresis in Sec. \ref{Sec:PH}.  Dynamical mechanism behind the pattern is discussed in Sec. \ref{Sec:DM}. Some other aspects such as the dimensionality and the types of perturbation are studied in Sec. \ref{Sec:OA}. Finally, we summarize our work in Sec. \ref{Sec:SD}. 

\section{Model description}\label{Sec:MD}
\subsection{Mean field treatment without space} \label{Sec:MF}

As in Ref. \cite{chen2016} we shall only consider the case of two infections A and B, each of SIS (Susceptible-Infected-Susceptible) type contagion dynamics. For a single infection of SIS type, host individuals can either be susceptible (S) or infected (I), the transmission happens via $S+I \rightarrow 2I$ and recovery by $I\rightarrow S$, with infection rate $\alpha$  and recovery rate $\beta$, respectively. The dynamics of SIS therefore captures a class of contagions that the recovered individuals carry no immunity and can be repeatedly infected during their lifespan, such as most seasonal flu. In a well-stirred population, one can write down the kinetic equations for $S(t)$ and $I(t)$, outbreak happens only if the so called basic reproduction number/ratio $R_0=\alpha/\beta>1$, and contagion-free otherwise.

When the SIS dynamics is generalized into the case of two agents (see Fig.~\ref{fig:1}a), a host could then be in one of four states $S$, $A$, $B$, $AB$, corresponding to being susceptible, infected with A only, with B only, and infected with both, respectively. In the transmission dynamics, we distinguish two infection rates: the initial rate $\alpha_A$ ($\alpha_B$) -- the rate that an agent A or B transmits to a susceptible $S$; and the secondary rate $\alpha_{AB}$ ($\alpha_{BA}$) -- the rate that a secondary agent transmits to a host that is already infected with A (B). For simplicity we assume uniform recovery rate $\beta$. With these, the full dynamics is described by
\small
\begin{eqnarray} \label{eq:coinfection}
    \dot{S} & = & -\alpha_A S(I_A+I_{AB}) - \alpha_B S(I_B+I_{AB}) + \beta(I_A+I_B), \nonumber \\
    \dot{I}_A & = & \alpha_A S(I_A+I_{AB}) - \alpha_{AB}I_{A}(I_B+I_{AB}) + \beta (I_{AB}-I_A), \nonumber \\
    \dot{I}_B & = & \alpha_B S(I_B+I_{AB}) - \alpha_{BA}I_{B}(I_A+I_{AB}) + \beta (I_{AB}-I_B), \nonumber \\
    \dot{I}_{AB} & = & \alpha_{AB} I_A(I_B+I_{AB}) + \alpha_{BA}I_B(I_A+I_{AB}) - 2\beta I_{AB}.
\end{eqnarray}
\normalsize
Here $S$, $I_A$, $I_B$, and $I_{AB}$ denote the density of individuals in state $S$, $A$, $B$, and $AB$, respectively. The precise meaning of $\alpha_{AB}$ is the infection rate that a host already infected with agent A to be further infected with B, and vice versa. One can then conveniently define the \emph{cooperativity coefficients} $C_A\!=\!\alpha_{AB}/\alpha_B$ ($C_B\!=\!\alpha_{BA}/\alpha_A$), measuring the infection of agent A (B) -- induced change in the secondary infection rate of the other. When the two agents cooperate, the secondary infection is easier, i.e. $C_{A,B}\!>\!1$; $C_{A,B}\!<\!1$ implies competitive contagions, such as the case of cross-immunity; and if $C_{A,B}\!=\!1$ the two agents are neutrally interacting, essentially decoupled in their contagion processes.  Without considering birth and death processes, the above four compartments are in conservation within the focal population, i.e. $S\!+\!I_A\!+\!I_B\!+\!I_{AB}\!=\!1$. In our study, we consider symmetrical parameterization for simplicity, i.e. $\alpha_A\!=\!\alpha_B\!=\!\alpha$ for the initial infection, $\alpha_{AB}\!=\!\alpha_{BA}=\alpha'$ for the secondary infection, which then implies $C_A\!=\!C_B\!=\!\alpha'/\alpha\!\equiv\!C$. 

In Ref. \cite{chen2016}, this mean field treatment has been systematically studied, the main findings are: for strong cooperation ($C\!>\!2$), the contagion shows backwards bifurcations \cite{martcheva2006}, i.e. first order dynamical phase transitions with two different thresholds in the baseline reproduction number $R_0$, one for outbreak at $1$, the other for eradication at $2\sqrt{C-1}/C\!<\!1$; for weakly cooperative or competitive scenarios, the contagion transition is qualitatively the same as the traditional single infection, manifested itself as a continuous outbreak transition. General asymmetrical parameters do not change the results qualitatively. For details we refer to Ref. \cite{chen2016}.

\begin{figure*}
\includegraphics[width=0.6\columnwidth]{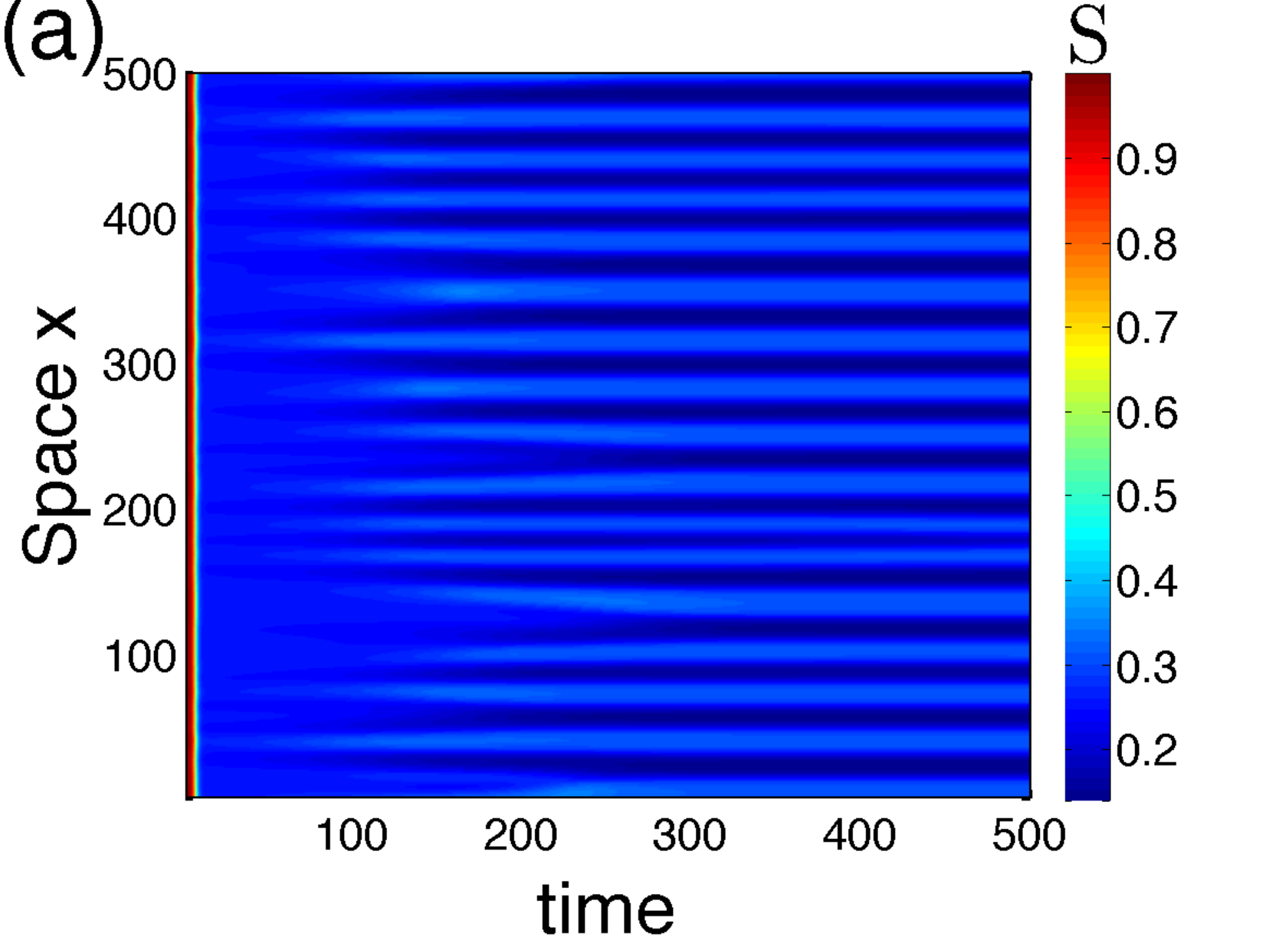}
\includegraphics[width=0.6\columnwidth]{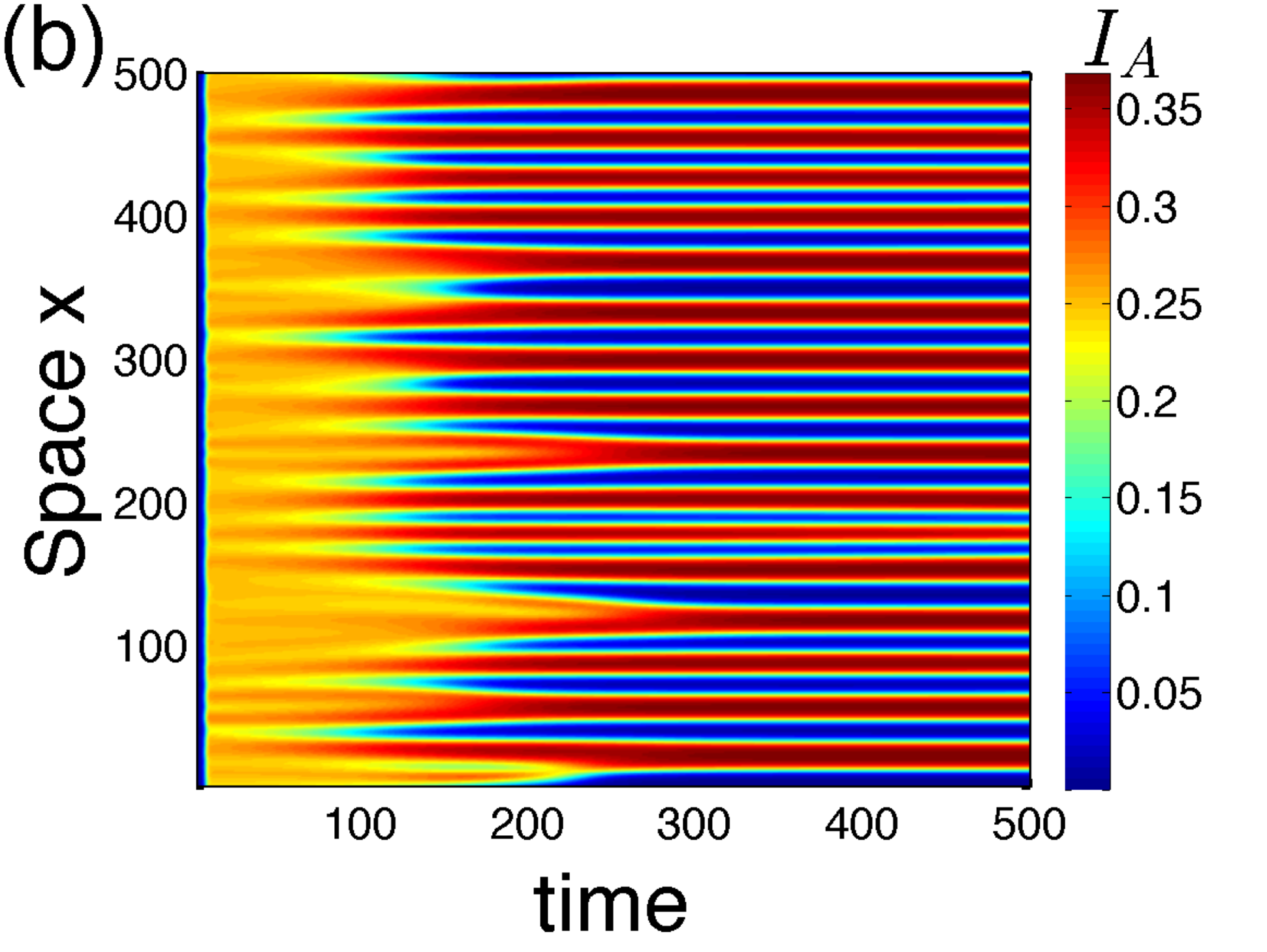}
\includegraphics[width=0.6\columnwidth]{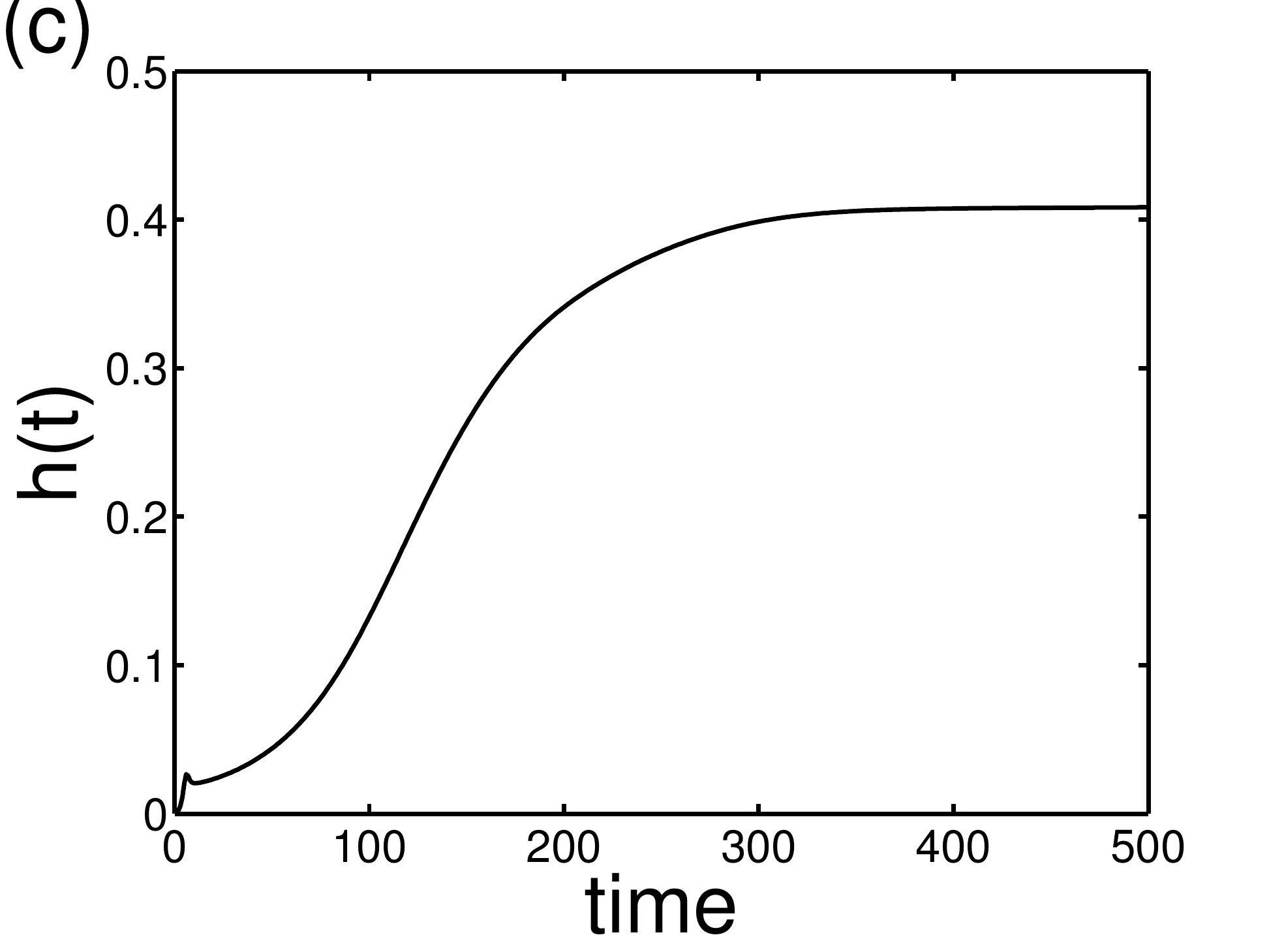}

\includegraphics[width=0.6\columnwidth]{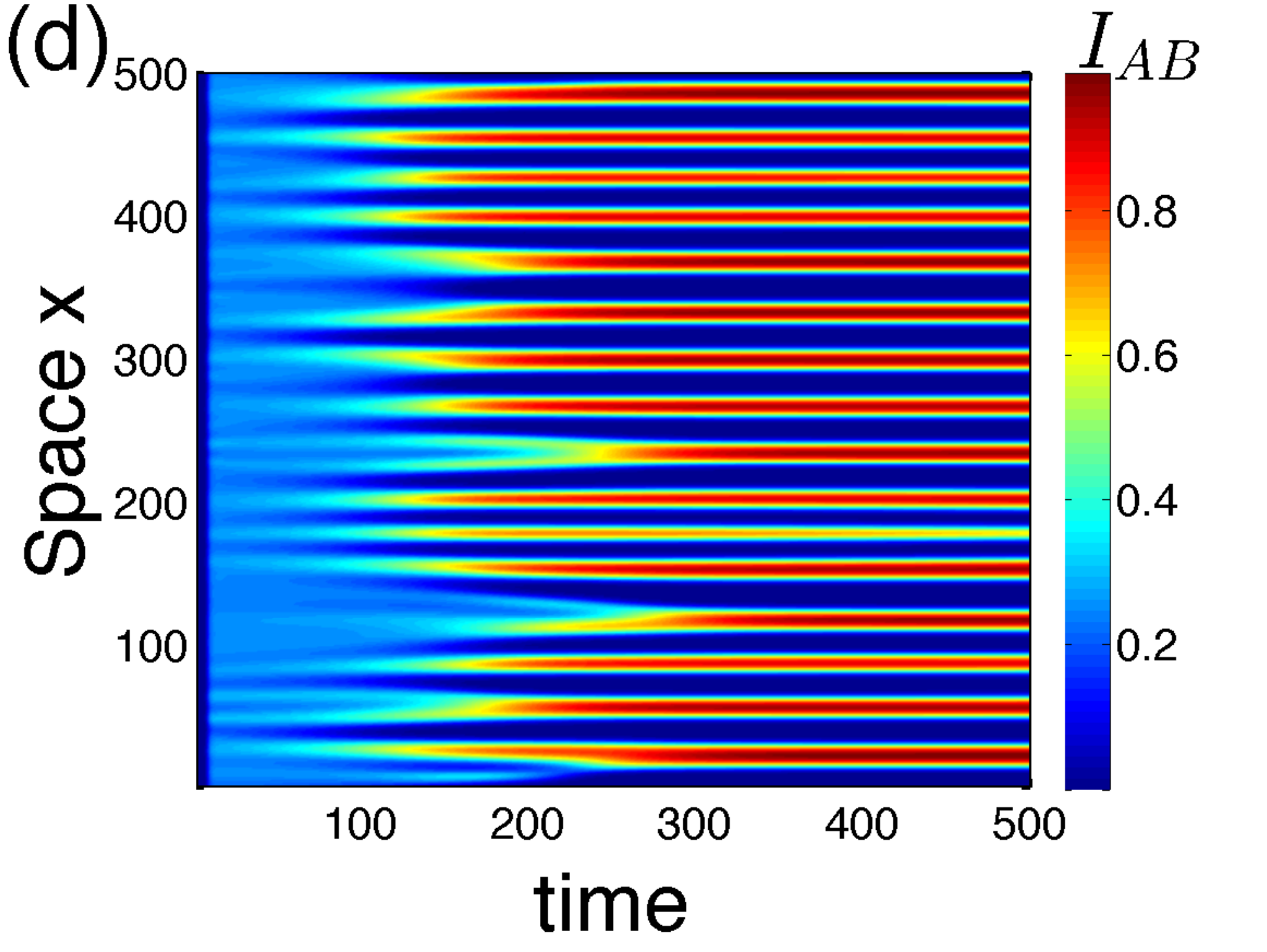}
\includegraphics[width=0.6\columnwidth]{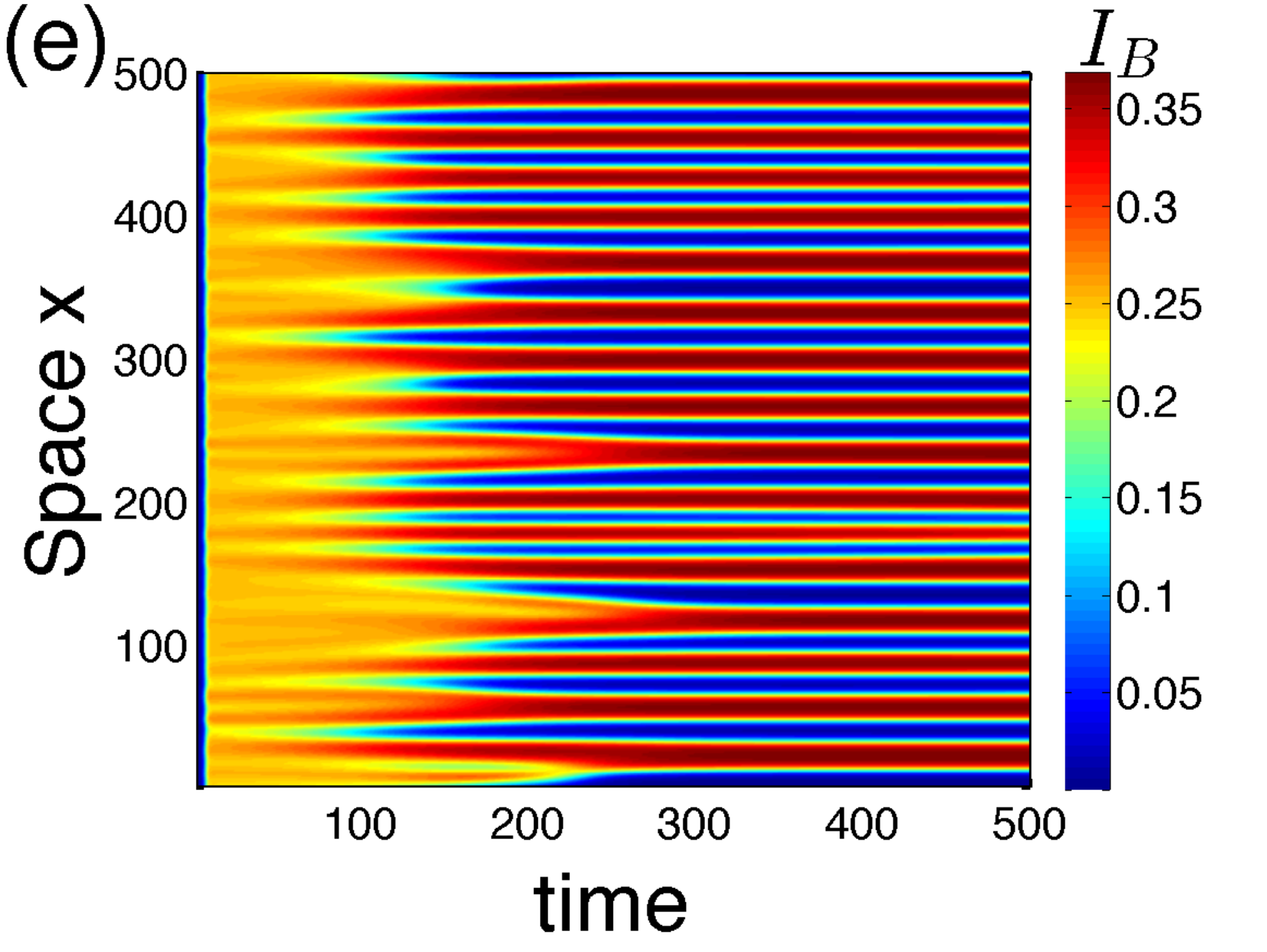}
\includegraphics[width=0.6\columnwidth]{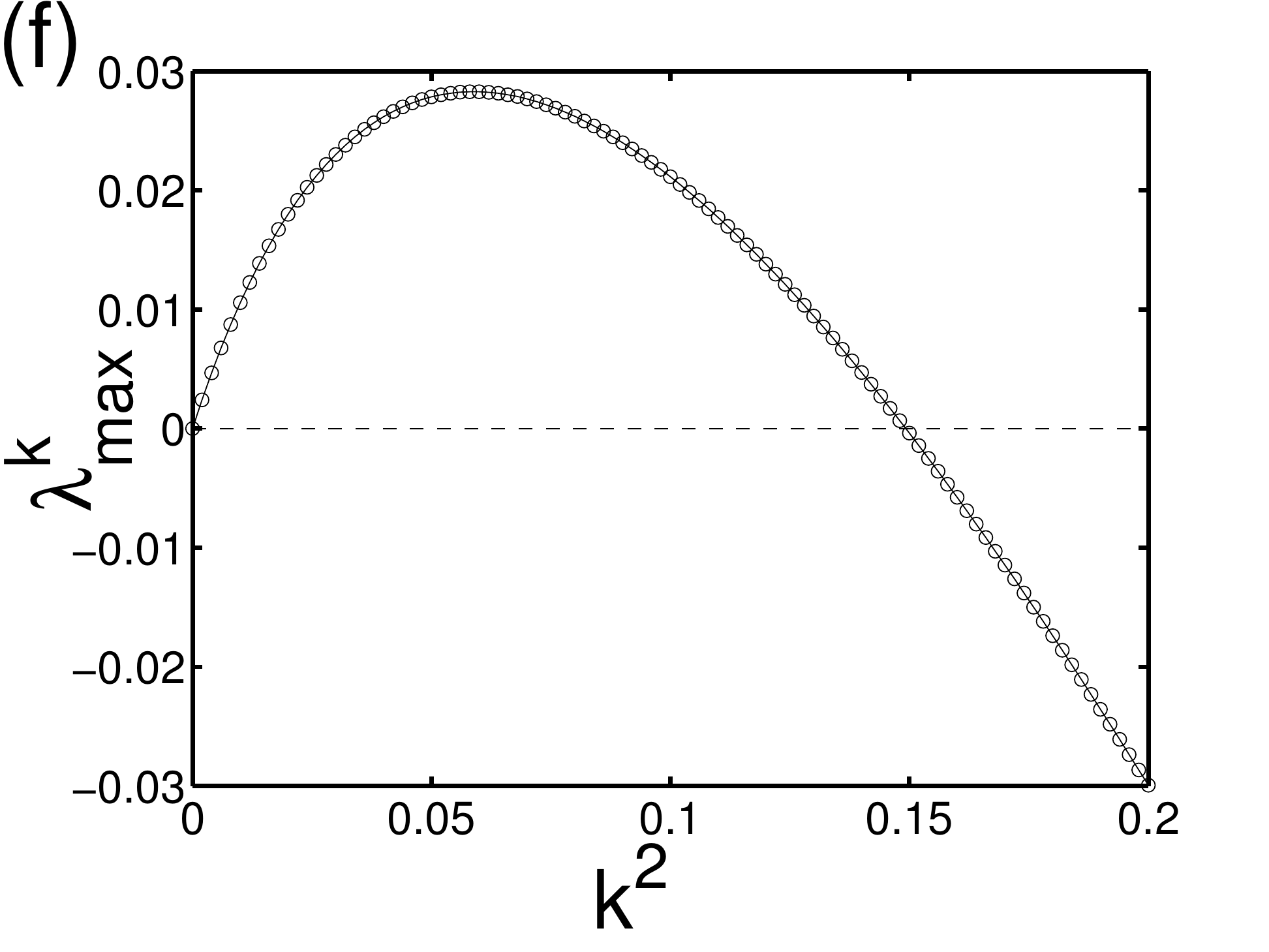}

\caption{\label{fig:2}\textbf{Emergence of infection patterns in $1d$ space.}
 Starting with perturbed contagion-free state, strongly and
weakly infected regions segregate from each other as time goes by (a,b,d,e). 
Note that, due to the symmetrical infection
parameters for the two infections, the resulting patterns of agent $A$ and
$B$ are also in symmetry as $t\rightarrow\infty$, even though their
initial conditions could be arbitrary.
The evolution of spatial heterogeneity in density $h(t)$ is shown in (c), 
showing that the heterogeneity gradually increases and then tends to saturation in the presented time window. 
 Subplot (f) shows that among
all Fourier modes, there are some unstable $\lambda_{max}^{k}\!>\!0$,
which trigger the spatial instability, in line with the patterns shown
here. Parameters: $R_{0}\!=\!2$, $C\!=\!1$, $D_{S}\!=\!10$, $D_{I}\!=\!1$. Periodic
boundary condition is used throughout the study.}
\end{figure*}

\subsection{Spatially interacting contagions} \label{Sec:SC}
When the spatial dimension is incorporated (Fig.~\ref{fig:1}b), the dynamics is conveniently described by the reaction-diffusion system \cite{murray2003}, which reads in $1d$ domain as
\begin{eqnarray} \label{eq:rd}
    \partial_t{S(x,t)} & = & f_S + D_S\partial_x^2 S, \nonumber \\
    \partial_t{I}_A(x,t) & = & f_A + D_A\partial_x^2 I_A, \nonumber \\
    \partial_t{I}_B(x,t) & = & f_B + D_B\partial_x^2 I_B, \nonumber \\
    \partial_t{I}_{AB}(x,t) & = & f_{AB} + D_{AB}\partial_x^2 I_{AB}.
\end{eqnarray}
The first terms in the rhs. $f_{S,A,B,AB}$ represent the intrinsic contagion dynamics, the same as the rhs. of Eq. (\ref{eq:coinfection}); the second terms capture the local mobility with that individuals potentially are capable of carrying the infected agents to their neighboring regions. $D_{S,A,B,AB}$ are the corresponding diffusion coefficients. The simplest case where $D_S\!=\!D_A\!=\!D_B\!=\!D_{AB}$ has been studied in \cite{chen2016}, mainly focusing on the properties of traveling waves. There, apart from the  classic scenario of forward traveling wave, backward propagation also emerges, together with the possibility of standing wave being expected. The later two new modes come from the competition between the reaction and diffusion in Eq. (\ref{eq:rd}).

In what follows, we are going to consider a more general setting, where the individuals' mobilities depend on their states; therefore they are not all identical any more. For the sake of simplicity, we only differ those infected from the susceptible, and do not further distinguish those partially infected and doubly infected, i.e. $D_A\!=\!D_B\!=\!D_{AB}\!=\!D_I\neq D_S$. In the context of epidemic spreading, those infected normally move less (e.g. staying home or hospital for recovery) than the healthy individuals, therefore we let $D_S\!>\!D_I\!=\!1$ if not stated otherwise. In Appendix A, the linearization analysis of Eq.(\ref{eq:rd}) is conducted, along with positive eigenvalues as the instability indicators for pattern to emerge. 

In numerical simulations, we define a spatial heterogeneity $h(t)$ to measure the emergence of pattern as following
\small
\begin{eqnarray}
h(t)=\sqrt{{\frac{{1}}{L_1}\intop_{0}^{L_1}\sum_{j=1}^{4}(X_{j}(x,t)-\langle X_{j}(t)\rangle)^{2}dx}}
\end{eqnarray}
\normalsize
 in $1d$ continuous space, or
\small
\begin{eqnarray}
h(t)=\sqrt{{\frac{{1}}{L_{1}L_{2}}\!\!\int_{0}^{L_{2}}\!\!\!\int_{0}^{L_{1}}\!\!\sum_{j=1}^{4}(X_{j}(x,y,t)\!\!-\!\!\langle X_{j}(t)\rangle)^{2}dxdy}}
\end{eqnarray}
\normalsize
in $2d$ continuous space. $L_{1,2}$ being the size of the domain
and $X_{1,2,3,4}=$$\left\{ S,I_{A},I_{B},I_{AB}\right\} $. $\langle X_{j}(t)\rangle$$ $
is the average density of each component over the whole domain. In our practice, we compute the heterogeneity according to
their discrete version
\small
\begin{eqnarray}
h(t)=\sqrt{{\frac{{1}}{N}\sum_{i=1}^{N}\sum_{j=1}^4(X_j^{i}(t)-\langle X_j(t)\rangle)^{2}}},
\end{eqnarray}
\normalsize
where $N$ is the number of local sites (including both $1d$ and $2d$
domains). By definition, a homogeneous solution (no pattern) means $h\rightarrow0$, and the heterogeneous case (pattern
emergence) results in $h>0$. Note that, in the spatial contagion of a single classic SIS agent, a well-known fact is that no positive eigenvalue is detected, the homogeneous state with $h(t\rightarrow\infty)=0$ is the only stable solution.

\section{Pattern Formation} \label{Sec:PF}

We start with $1d$ space, where we can see the spatial-temporal evolution of infection patterns. Figure~\ref{fig:2} shows an example in supercritical region ($R_0\!=\!2$), but without any cooperation or competition ($C\!=\!1$) at the moment, and the susceptible move faster than the infected $D_S\!=\!10$. The domain is initialized with infection-free state with tiny infected seeds, with periodic boundary.

As we can see, pattern emerges as the strongly and weakly infected regions are gradually formed and segregated.  A close comparison shows that the densities of four compartments are well correlated in the domain, where the density landscapes of $A$, $B$, and $AB$ overlap, and the component $S$ is opposite in the density as expected. In particular, the infection patterns of $A$ and $B$ are asymptotically identical $I_A(x,t)\!=\!I_B(x,t)$ when $t\rightarrow \infty$. For this reason, in the following we will adopt the overall density of agent A ($\rho_{_A}\!=\!I_A\!+\!I_{AB}$) as our observable to  illustrate the pattern, but bear in mind that the results apply exactly to the other agent B as well since $\rho_{_A}\!=\!\rho_{_B}$ ($\rho_{_B}\!=\!I_B\!+\!I_{AB}$) after transient. Note that, due to the difference in diffusion coefficients, the overall density of a given location is in general not conserved anymore, i.e. $S(x)\!+\!I_A(x)\!+\!I_B(x)\!+\!I_{AB}(x)\neq1$. The pattern formation process is captured in the increasing trend of spatial heterogeneity $h(t)$. By analyzing the eigenvalues of the linearized system, we indeed found that there is a positive eigenvalue region for some Fourier modes, which implies pattern formation and therefore is in line with our observations. The system exhibits a rich spectrum of interesting dynamical properties, which we will discuss in details in the following.

\subsection{Impact of contagion interaction}
\begin{figure*}
\includegraphics[width=0.6\columnwidth]{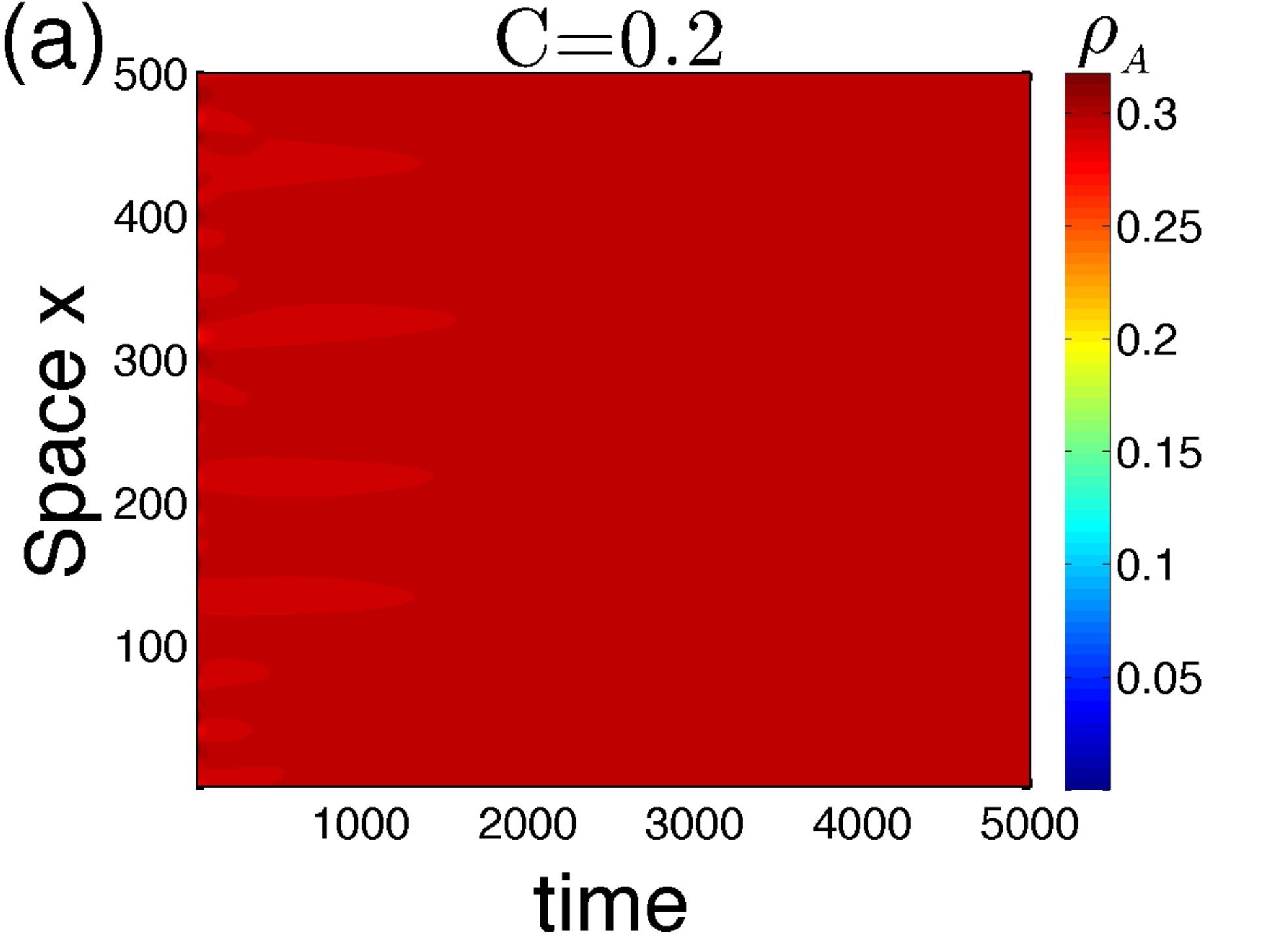}
\includegraphics[width=0.6\columnwidth]{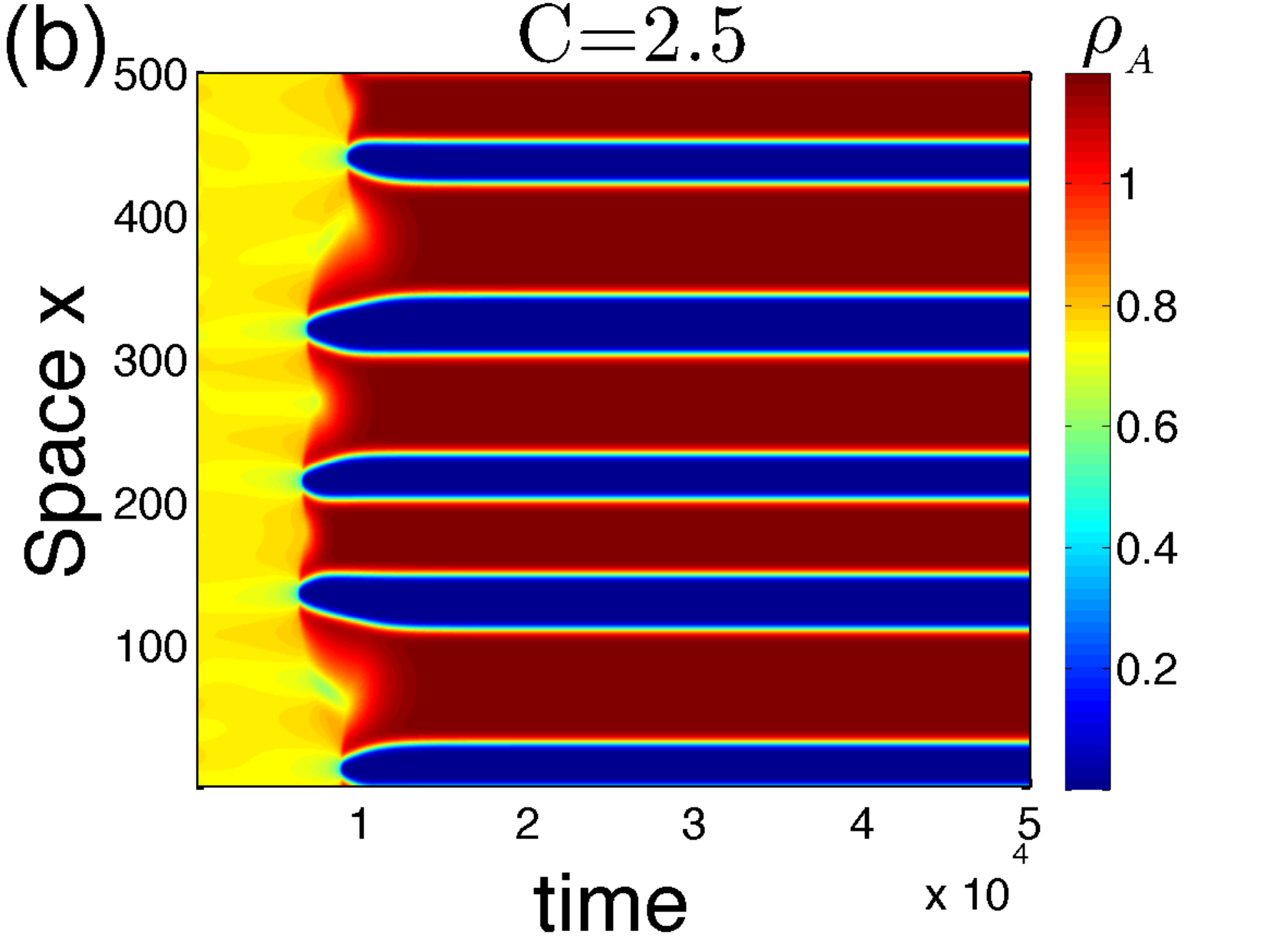}
\includegraphics[width=0.6\columnwidth]{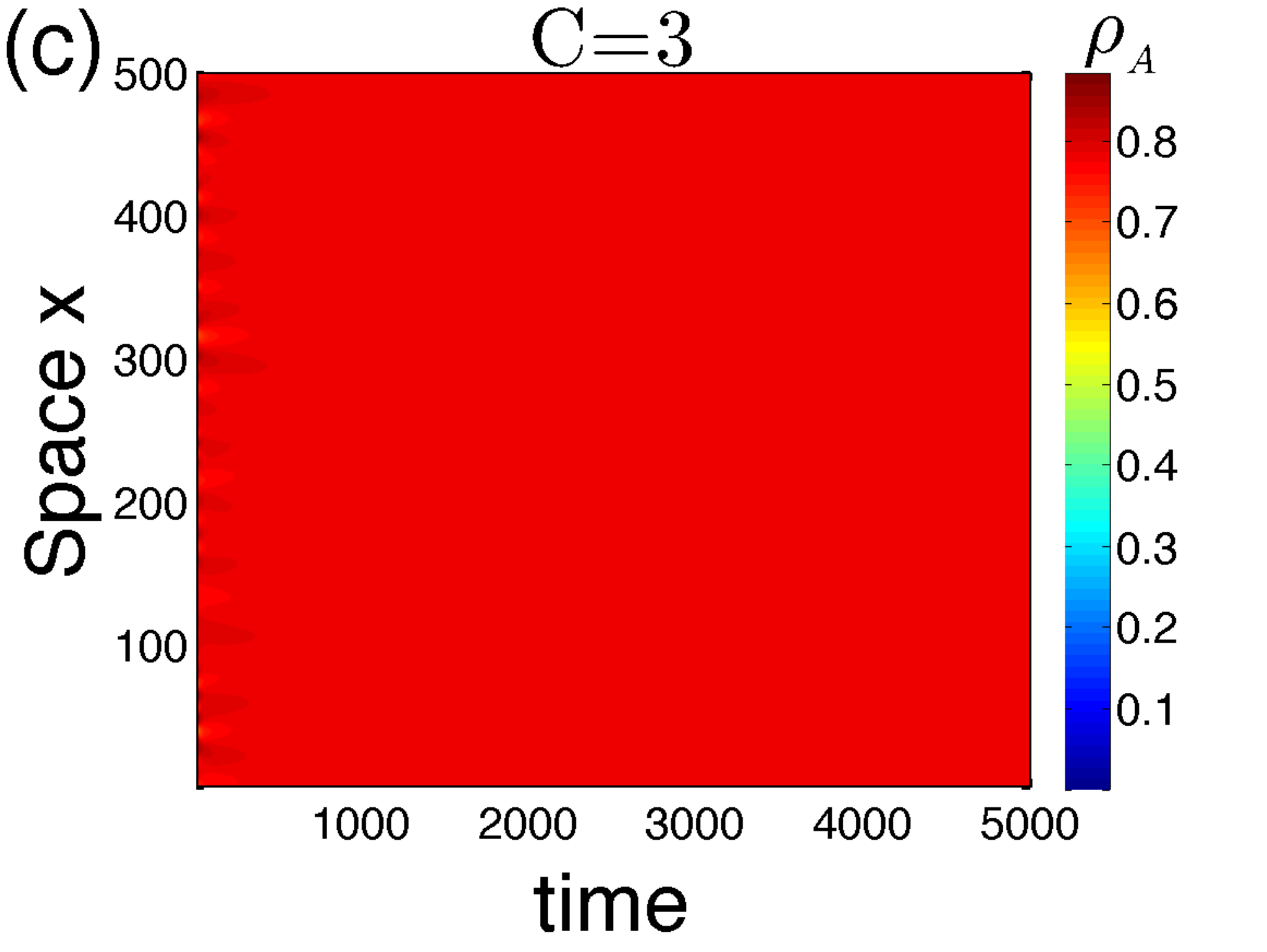}

\includegraphics[width=0.6\columnwidth]{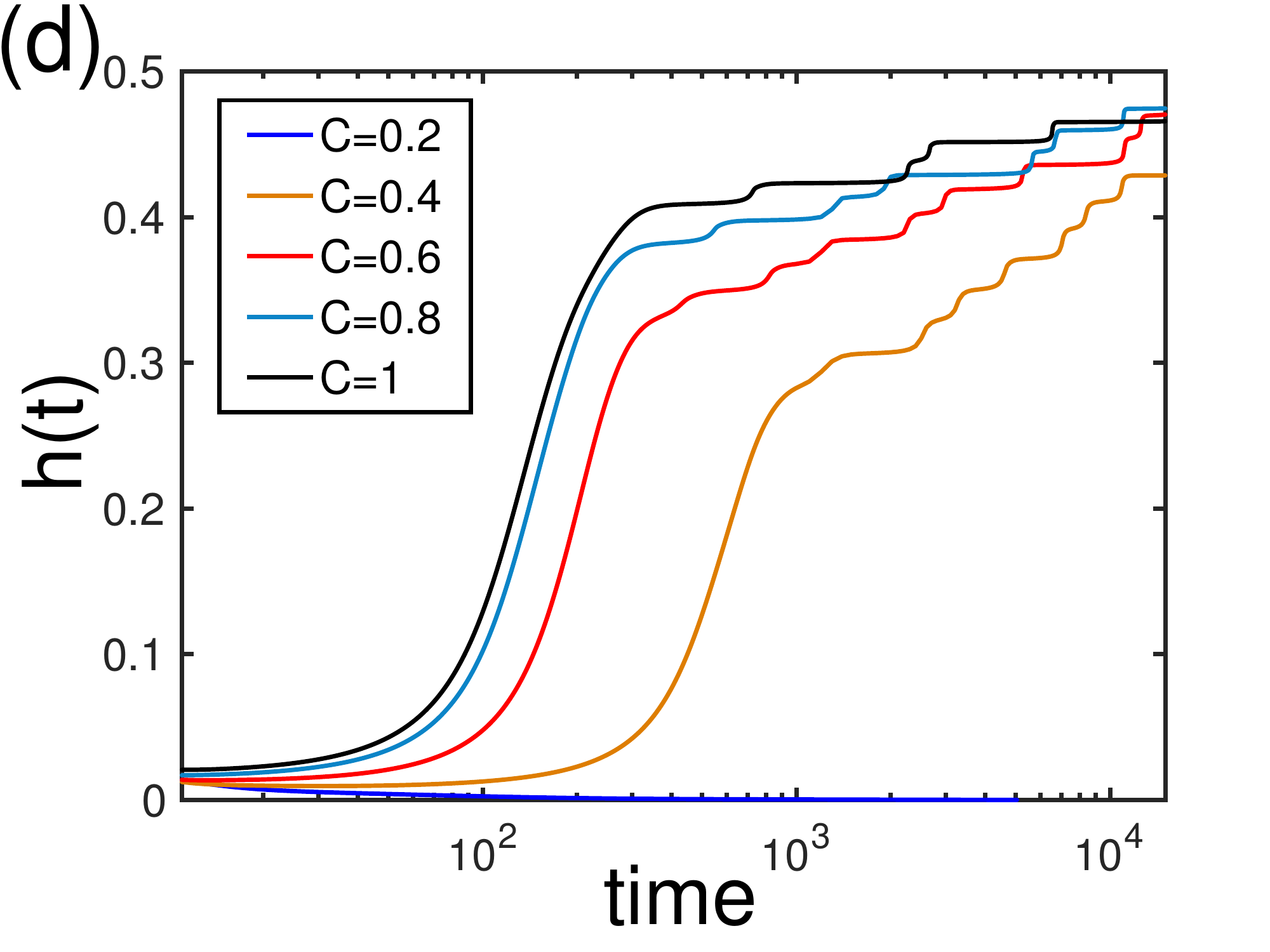}
\includegraphics[width=0.6\columnwidth]{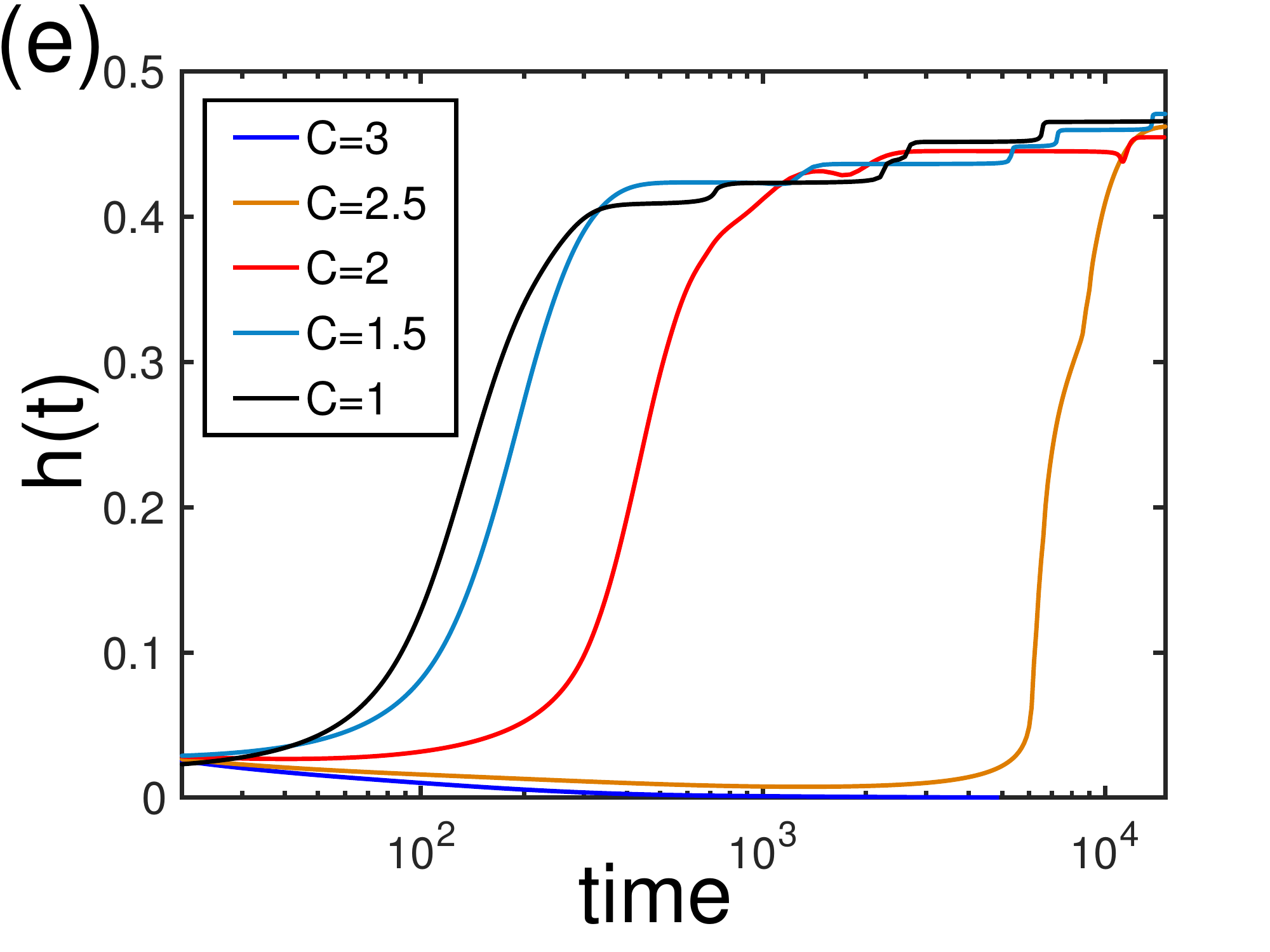}
\includegraphics[width=0.6\columnwidth]{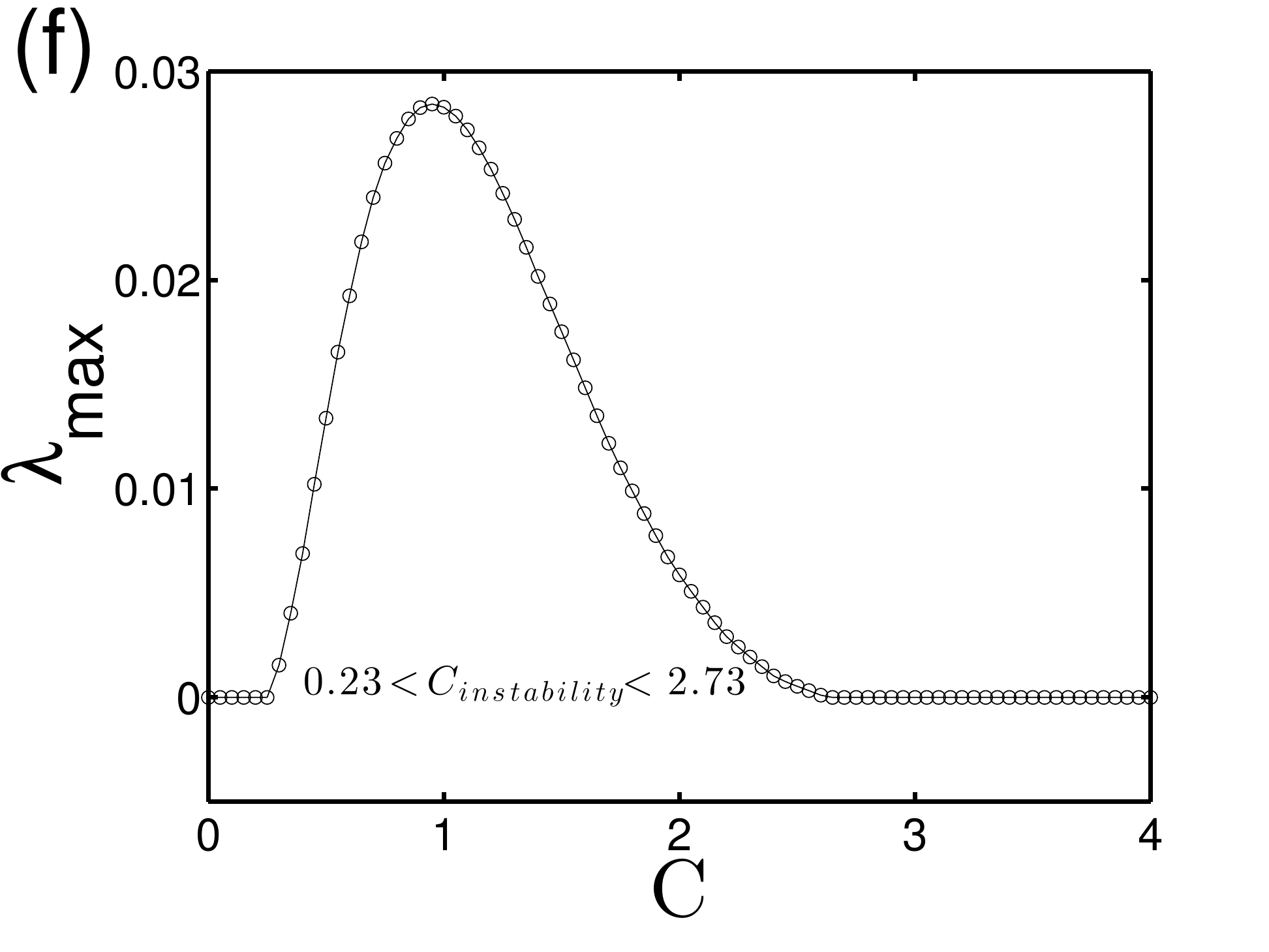}
\caption{\label{fig:3}\textbf{The impact of contagion interactions.} (Upper row) These plots show that strong competition
e.g. $C\!=\!0.2$ in (a), or strong cooperation like $C\!=\!3$ in (c) inhibits the emergence of pattern formation. The color-coded overall density of agent $A$ ($\rho_{_A}\!=\!I_A\!+\!I_{AB}$) is shown here. In particular, subplot (b) shows that a relatively strong interaction
($C\!=\!2.5$) delays the segregation process (notice the
time unit along $x$-axis). (Lower row) The plots of $h(t)$ for a couple
of interaction strengths, indicating that patterns formation is most
likely to happen in the case of $C\!\sim\!1$ (d and e), any deviation to a smaller or larger value will delay or just fail to have the formation process. Linearization analysis shows 
that pattern appears in a bounded range within $0.23\!<\!C_{instability}\!<\!2.73$, where $\lambda_{max}\!>\!0$
and the value peaks around $C\!=\!1$ (f). Parameters: $R_{0}\!=\!2$, $D_{S}\!=\!10$,
$D_{I}\!=\!1$.}
\end{figure*}
The first concern is the role of contagion interaction, because this is the main ingredient introduced here. One might think it is due to the contagion interaction that induces pattern formations. Strong cooperation or competition may be preferred. But this is not the case actually.

Figure~\ref{fig:3} shows that too competitive (small $C\!<\!1$) or too cooperative (large $C\!>\!1$) contagion interaction hinders the emergence of pattern (e.g. $C\!=\!0.2$ and $C\!=\!3$ in the upper row). More evidence is illustrated in the evolution of $h(t)$ for a couple of typical cases with different $C$ (Fig.~\ref{fig:3}d and~\ref{fig:3}e). Counterintuitively, for the cases of competitive interaction or cooperative type, the increasing trend of $h(t)$ become slower or just completely forbidden when the interaction $C$ deviates gradually from 1. This suggests the neutral scenario ($C\!=\!1$) is actually the most favored case for the emergence of pattern. Eigenvalue analysis clearly shows the impact of contagion interaction (Fig.~\ref{fig:3}f), where only a certain range of $C$ supports the pattern emergence, since positive eigenvalues only appear in a bounded parameter range, and the value peaks around $C\!=\!1$. While the sign of the largest eigenvalue indicates the possibility of pattern formation, the absolute value of positive ones determines the speed of segregation process. So for those with very small positive eigenvalues, the pattern formation takes a very long time, as shown in Fig.~\ref{fig:3}b.

Generally, contagion outbreak is the precondition of pattern formation; the presence of competition between two agents inhibits each other's outbreak,  therefore also suppresses pattern. This argument is in line with the observations that the competitive interaction deteriorates pattern dynamics. The above results, however, shows that cooperative interaction also impairs the emergence of pattern, which is counterintuitive, because a strong cooperation is always believed to facilitate the outbreak. A more confusing observation is that the neutral interaction case shows the optimal pattern scenario. This case is usually believed to be non-interacting, therefore the dynamics of the two agents are decoupled and no any pattern should be expected just as the single infection case.  This argument is actually not true because the two contagion processes are not completely decoupled when $C\!=\!1$. We will discuss these puzzles in the later part.

\subsection{Impact of mobilities}
\begin{figure}
\includegraphics[width=0.8\columnwidth]{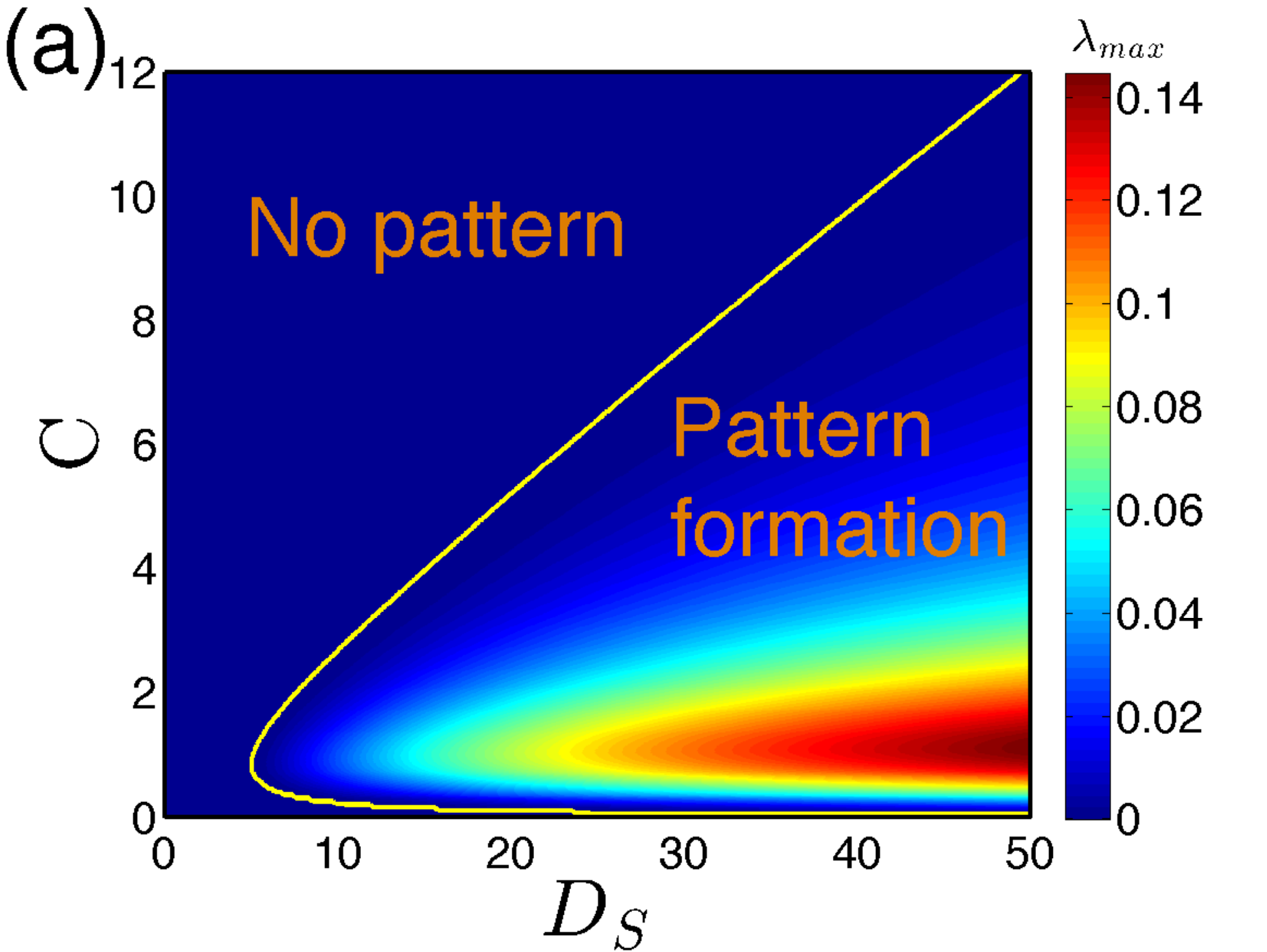}

\includegraphics[width=0.8\columnwidth]{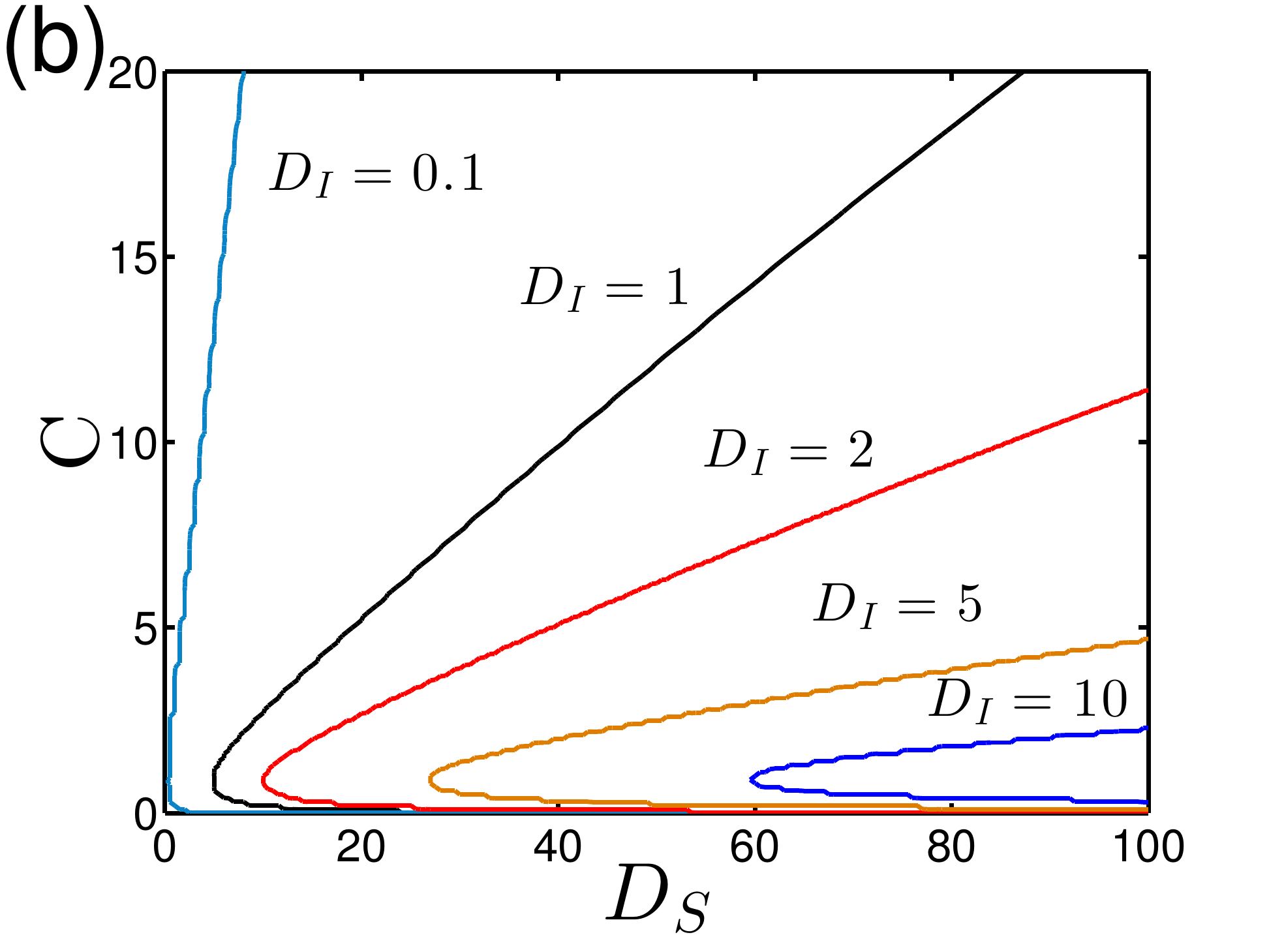}
\caption{\label{fig:4}\textbf{The impact of mobilities.} (a) The maximal eigenvalue $\lambda_{max}$
in $D_{S}-C$ parameter space, where the line separates the regions
with ($\lambda_{max}\!>\!0$) and without ($\lambda_{max}\!=\!0$) pattern formation. Here $D_I\!=\!1$ fixed. (b) Separating lines
for a couple of $D_{I}$, the left sides are parameter regions allowing pattern formation. 
Parameter: $R_0\!=\!2$.}
\end{figure}
Since a strong interaction is not a required ingredient for the pattern, we now turn to the impact of individual mobility -- the diffusion coefficients. Figure~\ref{fig:4}a shows for a given mobility of the infected ($D_I\!=\!1$), pattern tends to disappear when $C$ deviates from 1 or $D_S$ becomes smaller. The former observation is consistent with the above results that the neutral case ($C\!=\!1$) is optimal for pattern formation. The later observation in $D_S$ suggests that a higher mobility of the susceptible is beneficial to pattern emergence, and in principle large enough mobility in $D_S$ can always make the pattern formed no matter how strong interaction between the two agents (either very large or very small $C$).

Figure~\ref{fig:4}b shows the cases of different mobilities in the infected, where we can see that the higher mobility $D_I$ is, the smaller region is available for pattern formation. This means for those cases where the infected still move a lot, pattern is less likely to appear. Put together, pattern formation favors the condition when the susceptible diffuse a lot, and at the same time the infected move relatively less (i.e. large $D_S$, small $D_I$).

\subsection{Impact of the baseline reproduction ratio $R_0$}
\begin{figure}
\includegraphics[width=0.8\columnwidth]{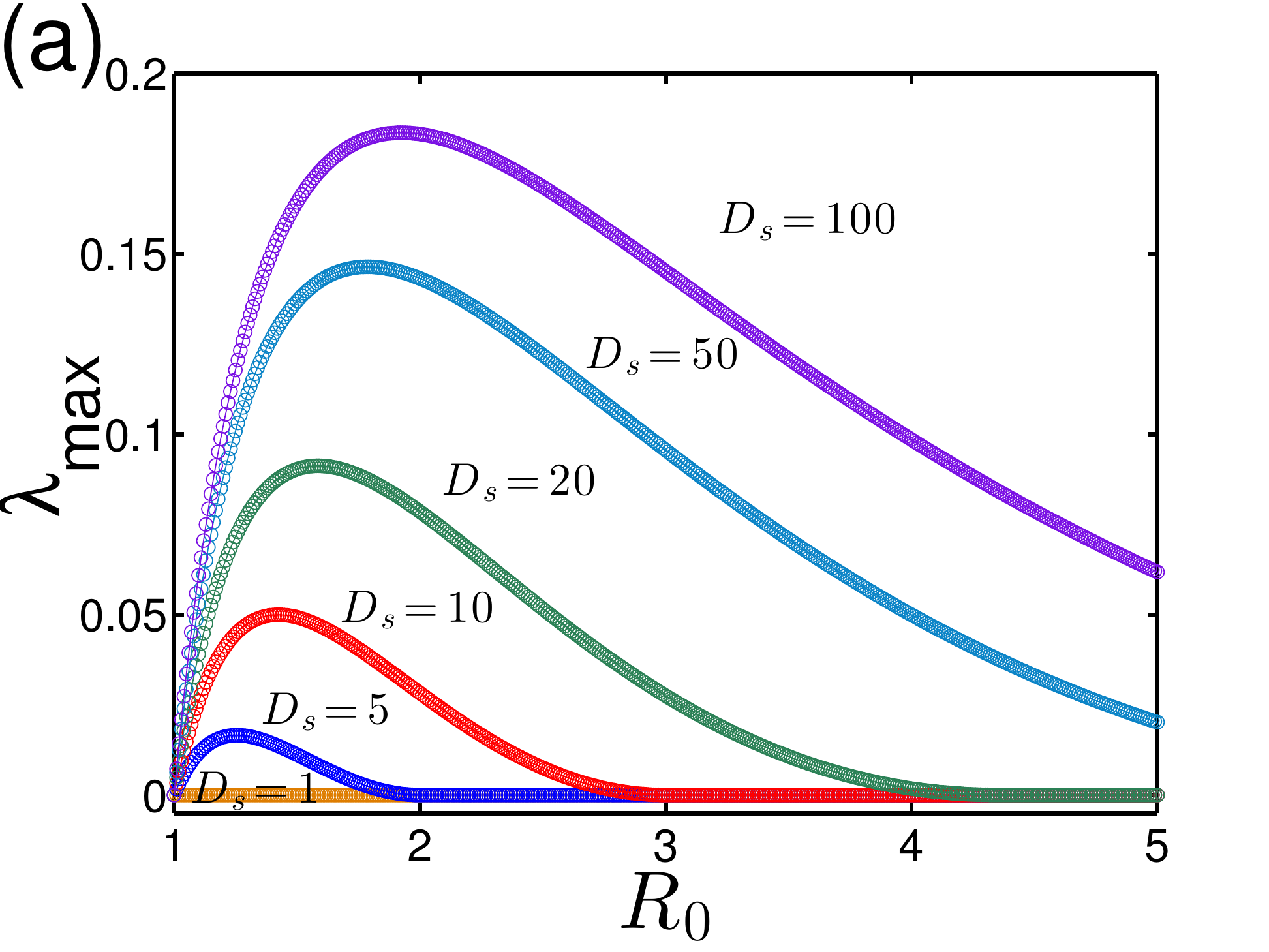}

\includegraphics[width=0.8\columnwidth]{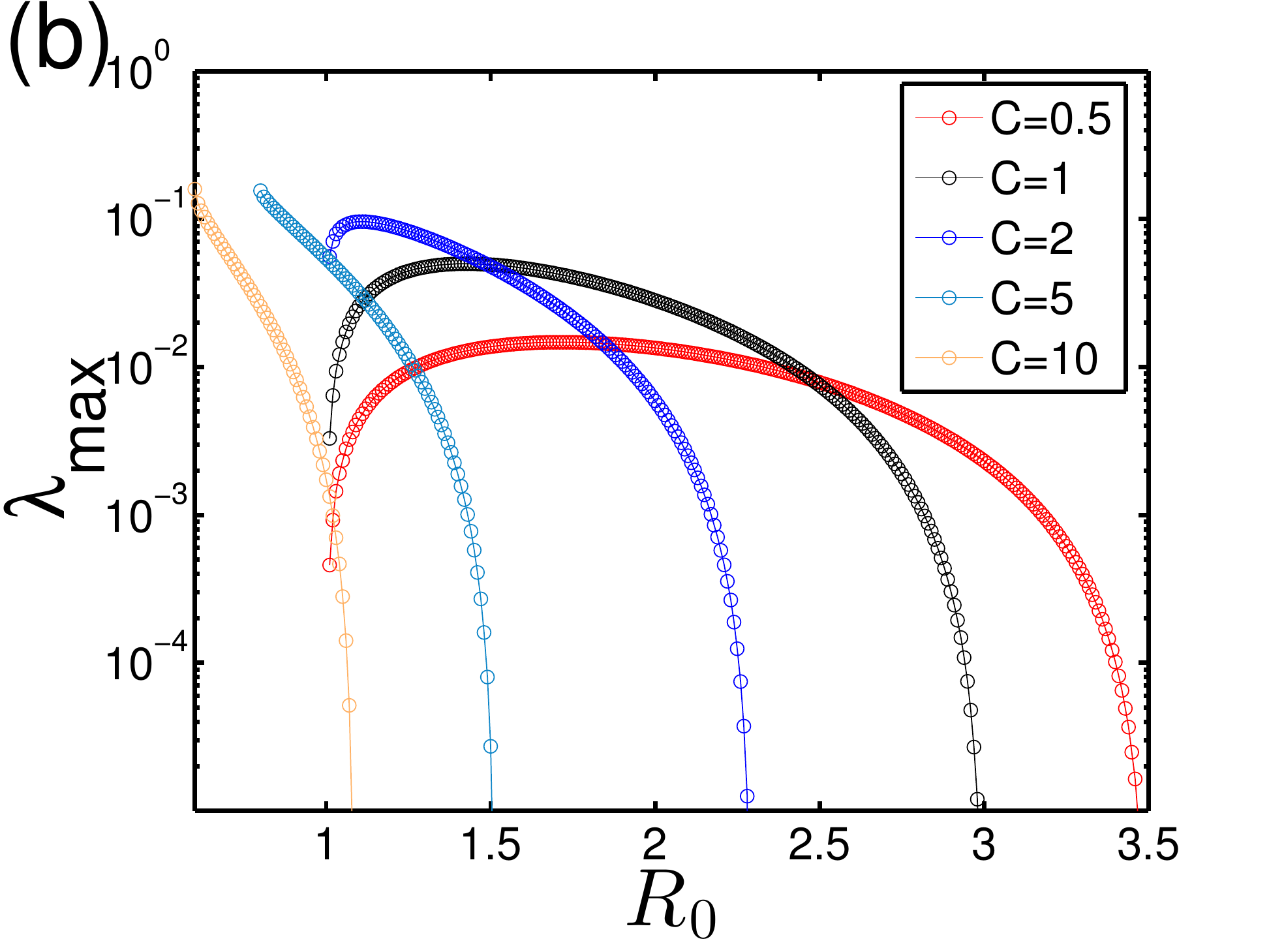}

\caption{\label{fig:5}\textbf{The impact of baseline reproduction ratio $R_0$.} 
(a) The maximal eigenvalue $\lambda_{max}$ versus $R_{0}$ for different $D_{S}$ by fixing $C\!=\!1$. 
The plot shows a bounded infection region in $R_0$ for pattern emergence, and that a higher mobility of the susceptible allows for a wider
pattern range of infection region. (b) The maximal eigenvalue $\lambda_{max}$ versus $R_{0}$ for different $C$ by fixing
$D_{S}\!=\!10$. These curves show there is an upper threshold for all cases, above
which the contagion pattern disappears. }
\end{figure}
As the traditionally primary control parameter in contagion dynamics, the reproduction number $R_0$ measures the baseline infection capability of a given agent. Figure~\ref{fig:5} shows the impact of $R_0$ for a couple of diffusion coefficients $D_S$ and contagion interaction levels $C$. 

Unlike outbreak transitions, there is an upper threshold for $R_0$ for pattern transitions as seen in both figures, above which the prevalence $\rho_{_{A,B}}$ can only present in the form of homogeneous distribution.  In Fig.~\ref{fig:5}a, the shared lower threshold ($R_0\!=\!1$) is due to the outbreak threshold for $C\!=\!1$ for all cases expect for $D_S\!=\!1\!=\!D_I$, which is too small to trigger a pattern. In the meantime, a higher diffusion in the susceptible lifts the upper threshold, though still cannot remove its presence. The region dependence on $C$ shows that a higher cooperation in the interaction considerably reduces the upper threshold (Fig.~\ref{fig:5}b), and a competitive interaction allows for patten with a larger reproduction number $R_0$. The lower threshold for the cooperative contagion, however, can be much reduced, as will shown in the next section. 

\section{Pattern Hysteresis} \label{Sec:PH}
\begin{figure*}
\includegraphics[width=0.7\columnwidth]{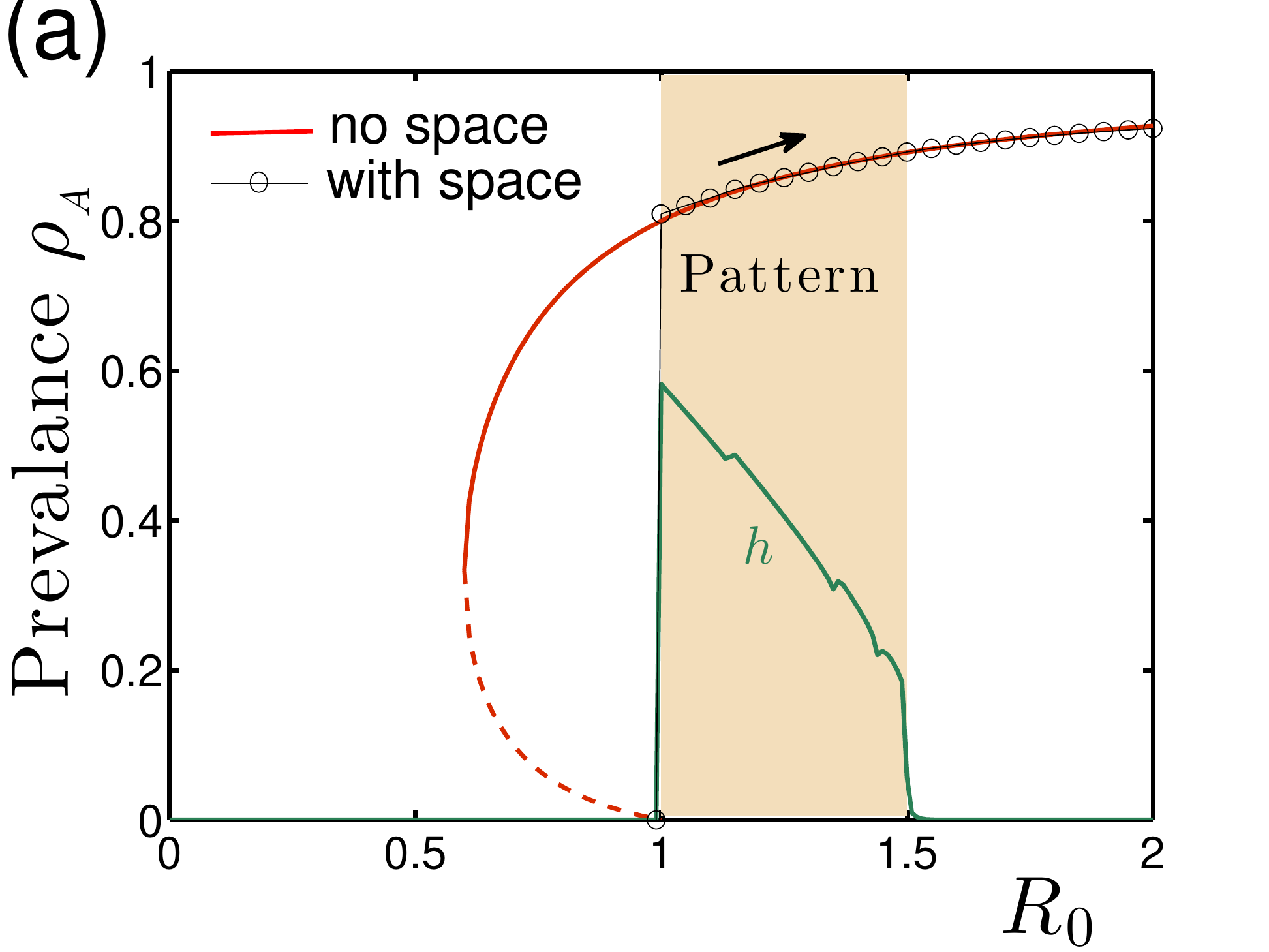}
\includegraphics[width=0.75\columnwidth]{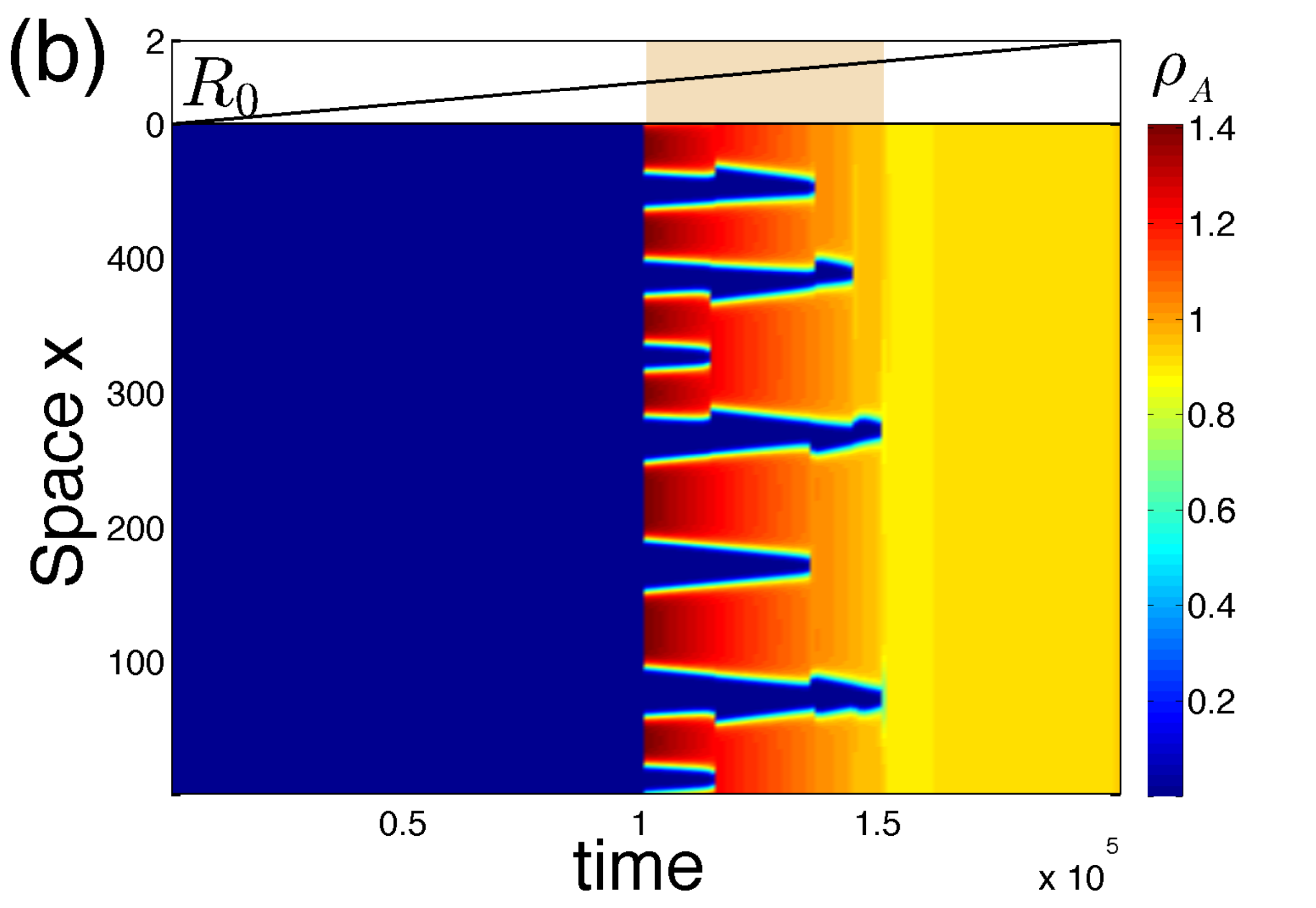}

\includegraphics[width=0.7\columnwidth]{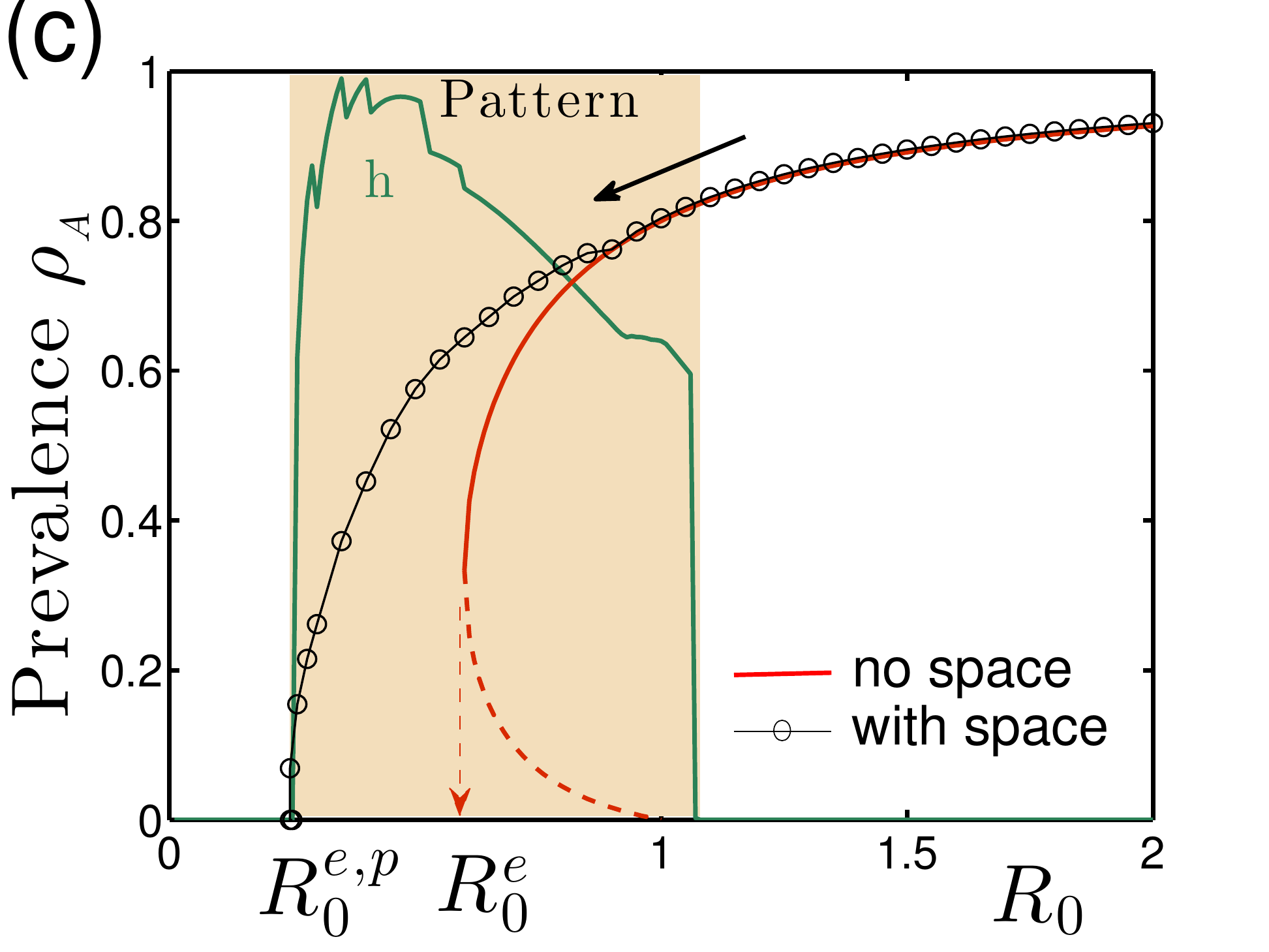}
\includegraphics[width=0.75\columnwidth]{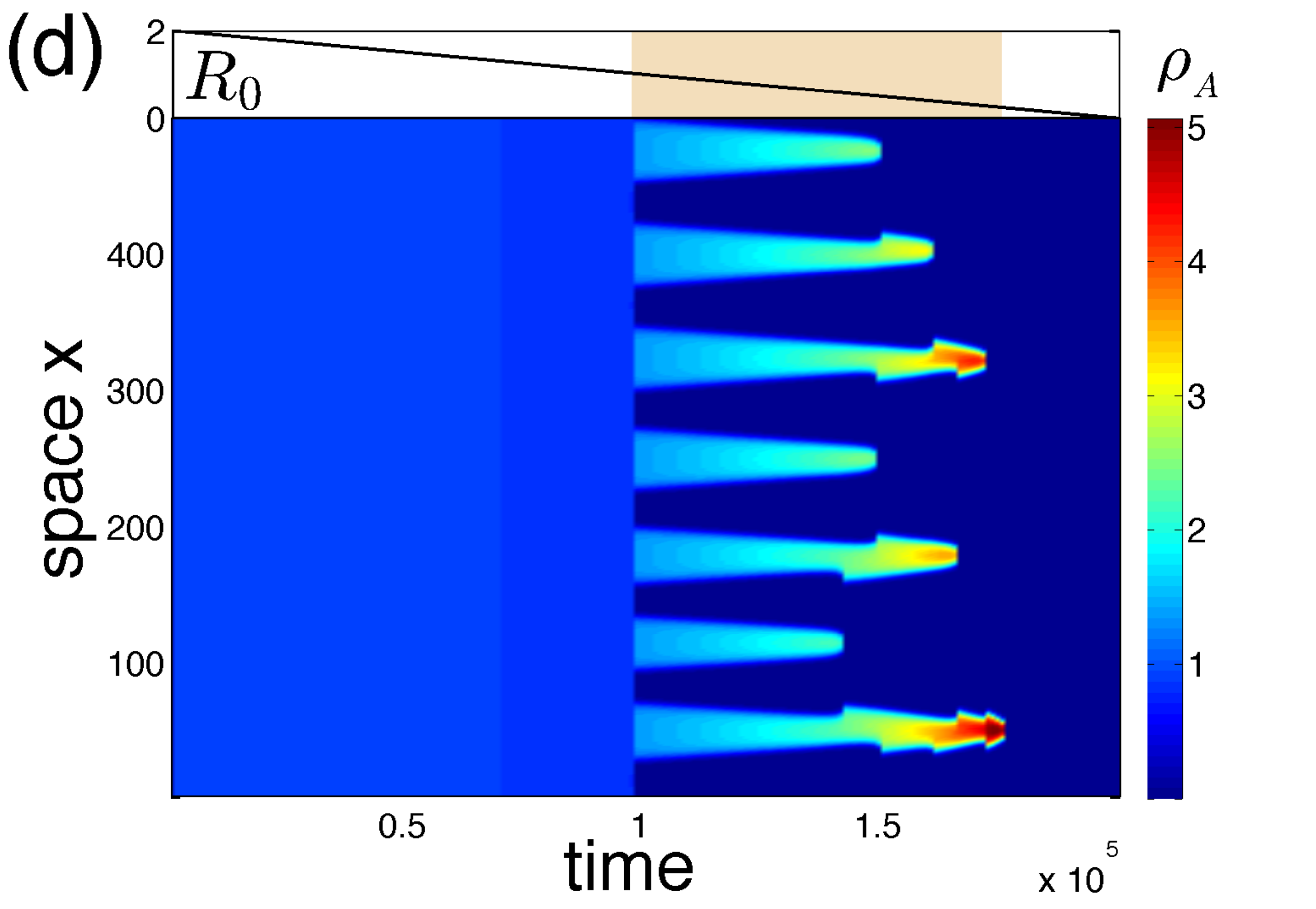}
\caption{\label{fig:6}\textbf{Pattern hysteresis.} 
Pattern hysteresis emerges when we increase the parameter $R_{0}$
(upper row) and then decrease it (lower row). (Left column -- (a) and (c))
The average prevalence as a function of $R_{0}$ with and without space in both directions, there the
pattern presents in the shaded region indicated by a nonzero
heterogeneity $h$. (Right column -- (b) and (d)) The corresponding spatial-temporal evolution of patterns. 
The parameter $R_0$ changing is slow enough (subplots at the top) to have a stable pattern for each parameter. 
Pattern hysteresis presents in the sense that in the two directions, the parameter
regions $R_0$ available for pattern are quite different (1 \textendash{} 1.5 for increasing in (a), 1.08 \textendash{}
0.25 for decreasing in (c)). The lower panels ((c) and (d)) show that the presence of pattern leads to
a much more stubborn eradication because $R_{0}^{e,p}\!<\!R_{0}^{e}$, close to zero,
a bad news for containment. Parameter:
$C\!=\!10$ with tiny conservative noise kept in the system.}
\end{figure*}
A main concern for contagion phenomena is the prevalence. Let's first focus on the cooperative cases, where the hysteresis is most prominent and we will discuss neutral and competitive cases later. Here we want to address the following question: \emph{Compared to the case without space (described by Eq.(\ref{eq:coinfection})), what is the impact of embedded space on the overall prevalence of outbreaks?} 

To this aim, we first slowly increase the baseline reproduction number $R_0$ of the noisy system from zero to trigger the outbreak, till a large value; then we decrease $R_0$ (\emph{e.g.} by vaccination programs in the context of infectious diseases) for the contagion eradication, and we examine the prevalence in the whole process. An interesting dynamical property we identified is hysteresis in pattern dynamics, as shown in Fig.~\ref{fig:6}. In the direction of increasing $R_0$ (Figs.~\ref{fig:6}a and~\ref{fig:6}b), outbreak transition remains the same for both cases with and without space, where the outbreak thresholds are identical both at $R_0\!\approx\!1$ and the two outbreaks share the same prevalence. Immediately after the outbreak, pattern is formed. Further increase in $R_0$ interestingly does not destroy the patterned infection as the theoretic prediction that when $R_0\!>\!1.08$ for $C\!=\!10$ (see Fig.~\ref{fig:5}) pattern should become unstable; Instead pattern disappears until $R_0\!>\!1.5$ as shown in $h(t)$. The presence or absence of pattern does not alter the overall outbreak size in this direction.

In the opposite direction (Figs.~\ref{fig:6}c and~\ref{fig:6}d), when we start with a large $R_0$, pattern is not permitted at the beginning. By slowly decreasing the reproduction number, pattern emerges until $R_0\!\approx\!1.08$ predicted by the eigenvalue analysis. The amazing phenomenon happens when we further decrease $R_0$ that the eradication of the two infections occurs not at $R^e_0$ but at a much smaller threshold $R^{e,p}_0$ (``p" indicates the presence of pattern). This means that a very stubborn prevalence presents in the system compared to the mean field case (without space) in this less infectious region. An intuitive explanation can be found in Fig.~\ref{fig:6}d, where the coinfected individuals are now clustered in a few spatial spots in quite high density, there the two agents support each other making their survival at very small $R_0$. This clustering behavior is crucial for the stubborn survival, and the spatial embedding as a new dimension provides such a possibility. 

Taken together, by varying $R_0$ differently, the parameter regions for pattern formation are quite different, only sharing little overlap. This process is reminiscent of  hysteresis in statistical physics, and we term the phenomenon as \emph{pattern hysteresis}. In the standard hysteresis like in the first-order phase transitions, there is a shared bistable region bounded by two transition points, the hysteresis is defined by two thresholding behaviors. In pattern hysteresis, however, there is only a tiny shared region with quite different transition points, the pattern dynamics occurs in different regions when moving along different directions. Pattern hysteresis leads to a stubborn prevalence for some locations, and to eradicate their infections, an unusually much effort is required.

\begin{figure*}
\includegraphics[width=0.6\columnwidth]{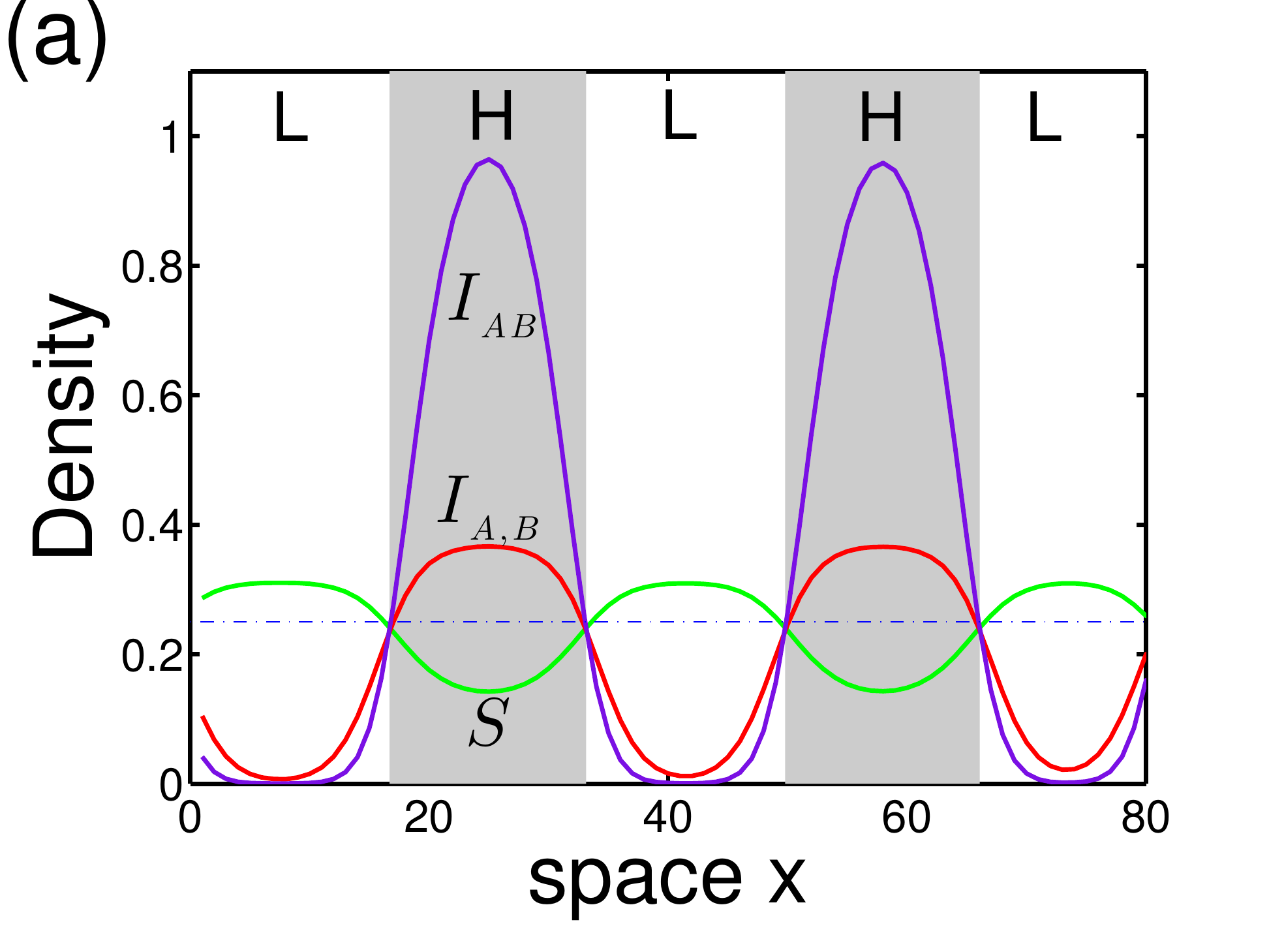}
\includegraphics[width=0.6\columnwidth]{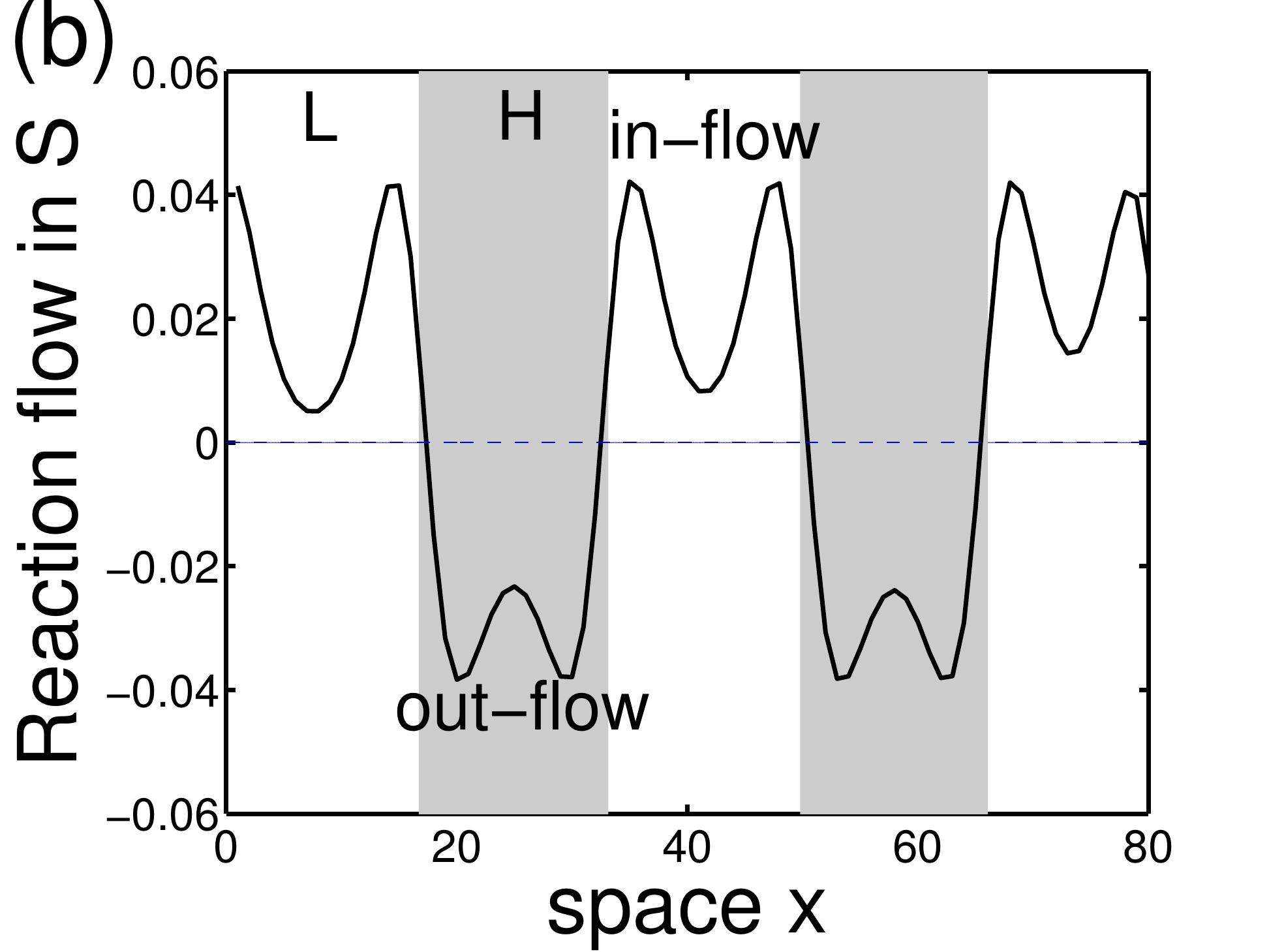}
\includegraphics[width=0.6\columnwidth]{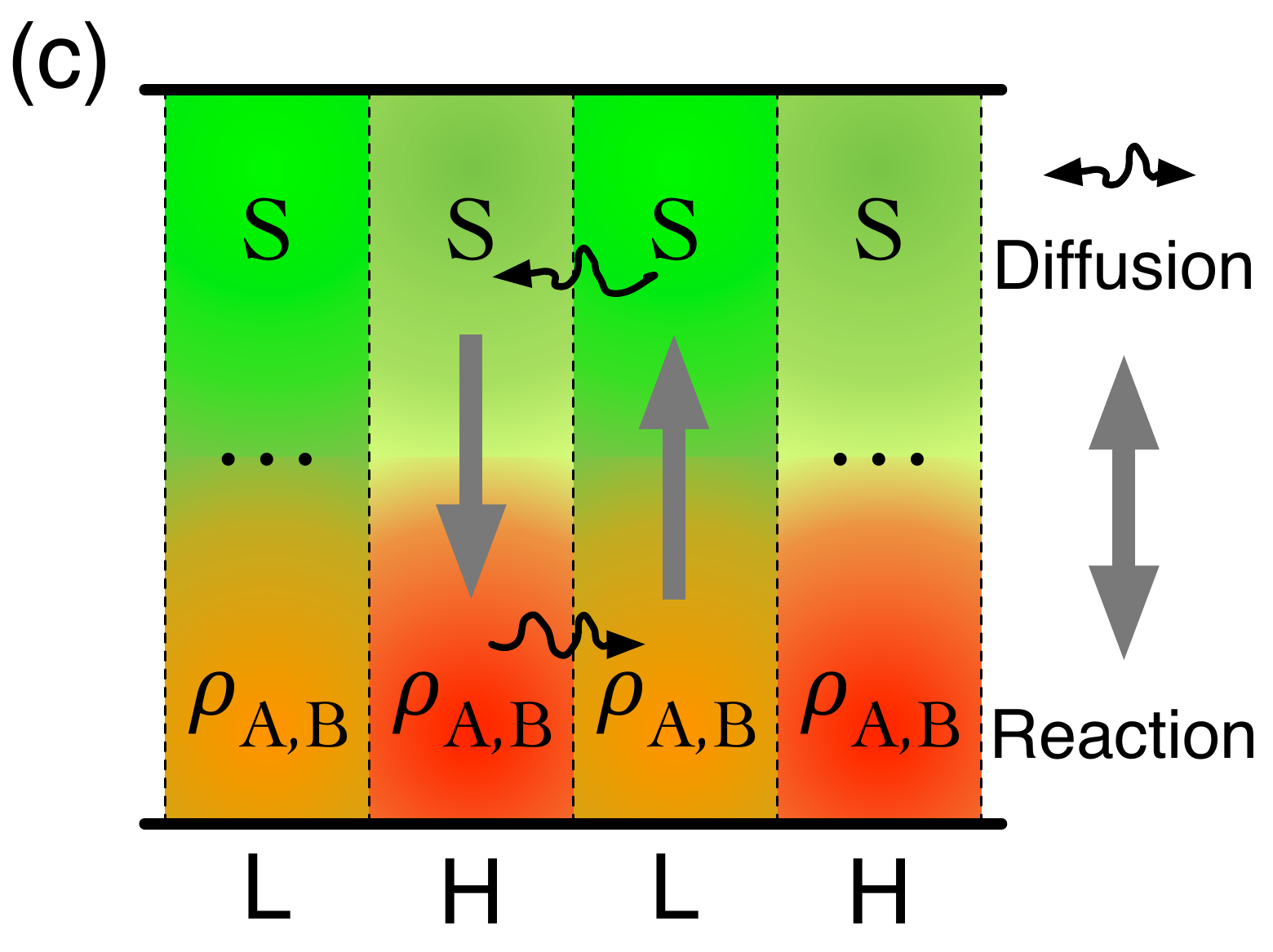}
\caption{\label{fig:7}\textbf{Dynamical mechanism.}  
(a)
Density profile of the four compartments in $1d$ domain. There, the high (H) and low (L) infected density regions are formed in sequence. Interestingly, the density distribution of $S$ is inverse compared to all other compartments. The dotted-dashed line is the mean field value (\emph{i.e.} without space) of the susceptible for reference.
(b)
The contribution of reaction in the dynamics of $S$ (\emph{i.e.} $f_S$ in Eq. (\ref{eq:rd})). Positive reaction flow (in-flow) means that the reaction tends to increase the density of $S$, whereas negative flow (out-flow) is to increase the infected by the consumption of $S$ instead.
(c) 
The scheme of dynamical flows between two neighboring regions: the net reaction flow within H regions is from the susceptible $S$ of a low density to the infected $\rho_{_{A,B}}$ of a high density to be even higher and is reversed in L regions -- a dynamical \emph{rich-get-richer} process (thick vertical arrows); Diffusion between neighboring regions is always from high density to low density as indicated by the curved arrows. In such a way, the four coarsen-grained components form a stable dynamical loop, leading to the pattern formation. Same parameters as in Fig.~\ref{fig:2}, and profiles in (a) are plotted after 1000 time units evolution. 
}
\end{figure*}

Unexpectedly, even for neutral ($C\!=\!1$) and competitive ($C\!<\!1$) contagions, the hysteresis still presents if a pattern is formed. For example, we computed the cases with $C\!=\!1$ and $0.5$ by slowly decreasing $R_0$, and found that the eradication threshold $R^e_0$ is below 0.7 and 0.9 respectively (data are not shown, other parameters are the same as in Fig.~\ref{fig:6}), smaller than the case without space, where $R^e_0\!=\!1$ at transition point. Due to the hysteresis-induced deviation from the mean-field value of prevalence, the computation of eigenvalues is not exact when without considering moving direction and only using the mean field prevalence (see Appendix A). True pattern regions with hysteresis are larger than the one computed from the fixed points in the mean field treatment, like Fig.~\ref{fig:5}.

\section{Dynamical mechanism} \label{Sec:DM}

Till now, we see the emergence of pattern and provide a theoretic analysis by computing the eigenvalues of the linearized system. But, still there is a lack of mechanism analysis, from which we may build an intuition in the understanding why these happen.

For this aim, we first plot the density distribution of all four components after a transient, see Fig.~\ref{fig:7}(a), from which we can read some important clues. We can see the densities of different components are now segregated within high- and low-density regions. For the convenience of discussion, we label the highly infected regions by H, and the lowly infected ones by L. The H regions are also of high overall density i.e. $S\!+\!I_A\!+\!I_B\!+\!I_{AB}$. But in these H regions the density of $S$ is instead lower than those in L regions. So the densities L/H distributions for $S$ and for those infected are reversal. To understand how these density profiles come to be possible, we next look into the contribution (\emph{i.e.} net flow) from the two different dynamical parts in the system. We found a \emph{rich-get-richer} aggregation process behind (Fig.~\ref{fig:7}(b)): In L regions, where the density of $S$ is higher than the neighboring regions, the reaction instead further increases its density to be even higher; a similar process happens also in H regions, where the high density of infected gets higher by the underlining reaction process.  Diffusion processes, however, always present to dilute any component of high density into its neighborhood, to counterbalance its accumulation. Therefore, the contributions from reaction and diffusion are just opposite.

\begin{figure*}
\includegraphics[width=0.65\columnwidth]{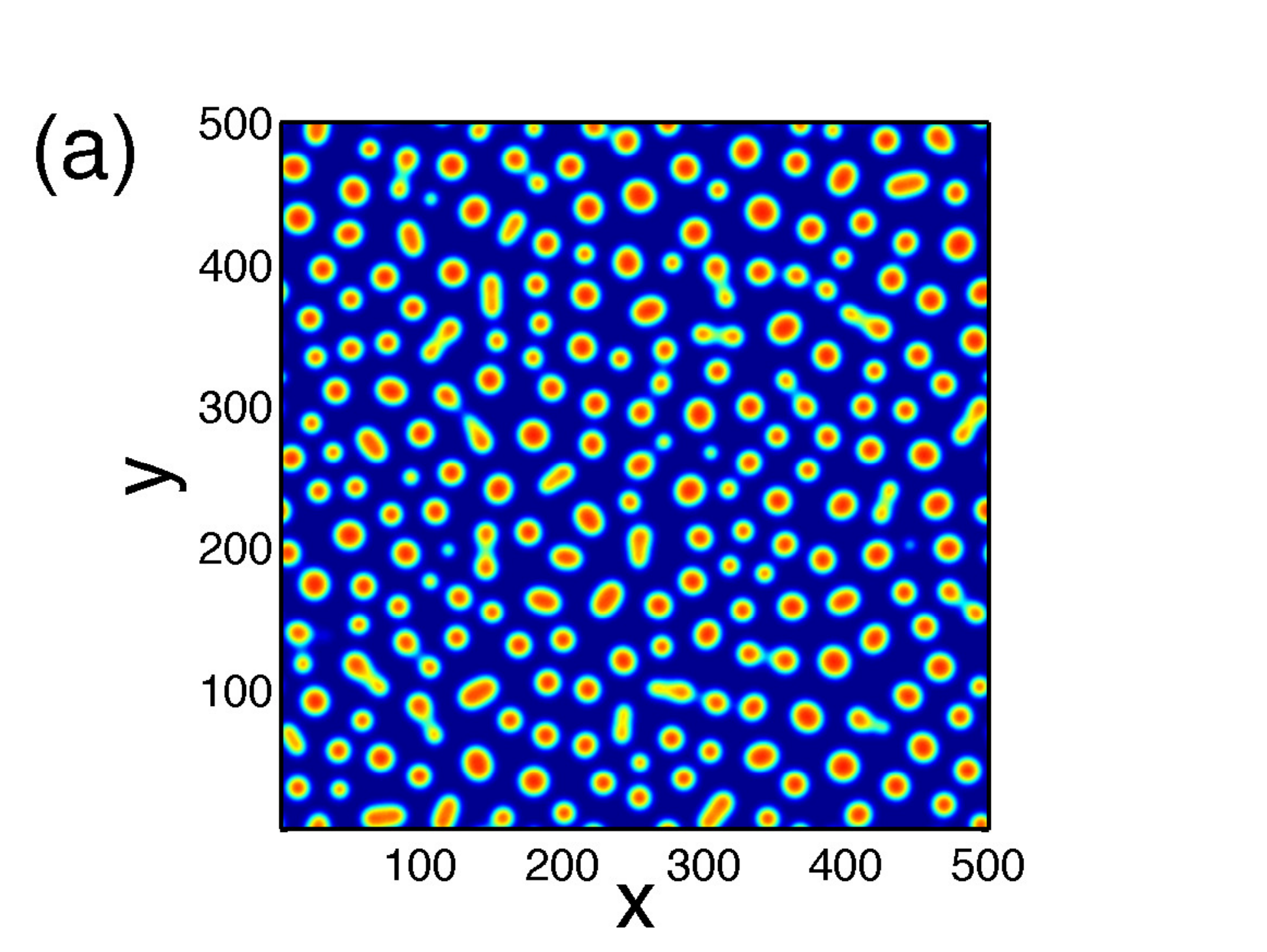}
\includegraphics[width=0.65\columnwidth]{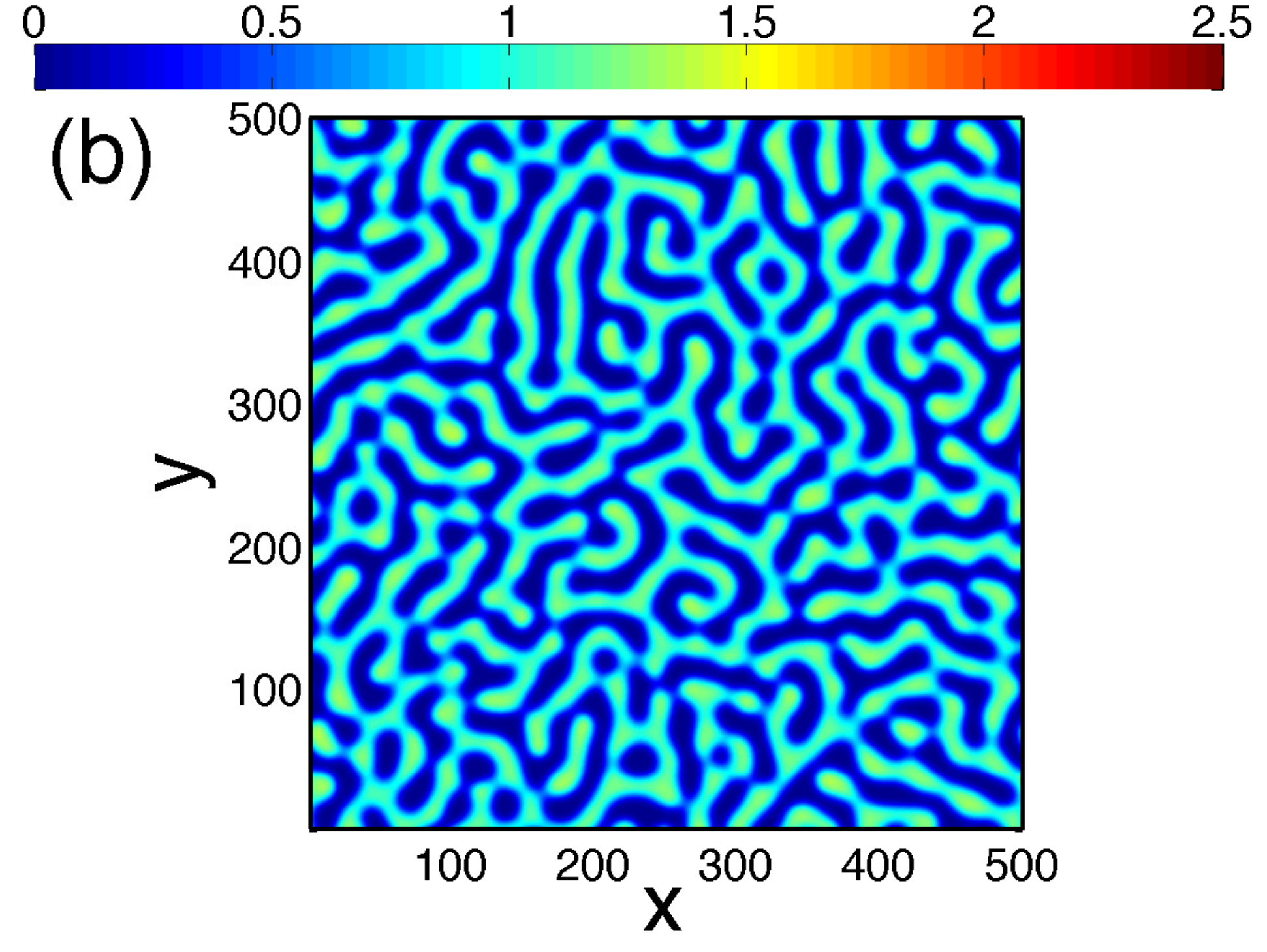}
\includegraphics[width=0.65\columnwidth]{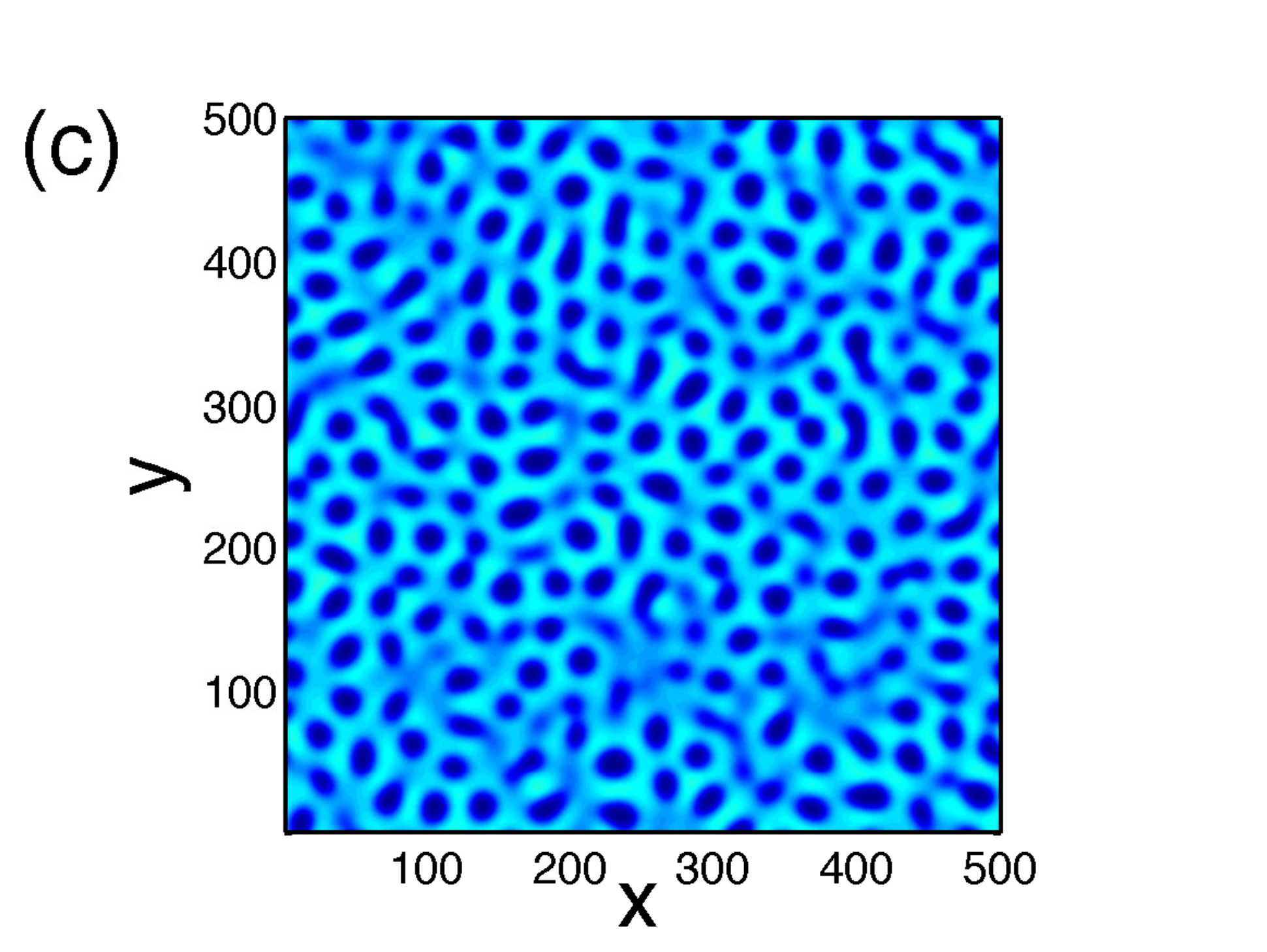}
\caption{\label{fig:8}\textbf{Typical patterns in $2d$ space.} (a) Spot
pattern for $R_{0}\!=\!1.5$ , (b) strip pattern $R_{0}\!=\!2.2$, (c)
spot pattern for $R_{0}\!=\!2.5$. The difference in the two spotted patterns is that the highly infected regions is the minority in (a) while the situation is reversed in (c). All patterns are shown starting from random
initial conditions and after 500 time units. Other parameters: $C\!=\!1$,
$D_{S}\!=\!10$.}
\end{figure*}

This key mechanism is summarized in Fig.~\ref{fig:7}(c). There the reactions happen within each local sites for aggregation, and diffusion occurs between neighboring sites for dilution. With these two sorts of dynamical flow, two neighboring sites, one with high density of infected (together with a low density of $S$) the other with low density of infected (with a high density of $S$), could then form a sustainable dynamical loop, supporting the pattern formation in spatially extended context. The stability of this dynamical loop depends on the interaction level $C$, a too small or too large value is found to break its sustainability. A too small $C$ means the termination in the reaction from other components to the coinfected $\rho_{_{AB}}$, this leads to two decoupled single-infection processes, $S\rightarrow A$ and $S\rightarrow B$. The local dynamics of rich-get-richer aggregation is then broken, no pattern expected. A too large value, on the contrary, leads to a dominating fraction in $\rho_{_{AB}}$, while other components especially the density of the susceptible becomes quite low in H regions. Under this circumstance, the inadequate supply of $S$ for the reaction to produce $\rho_{_{A,B}}$ makes the loop collapse, the pattern thus also fails to exist and homogeneous state becomes stable instead. For this strong cooperation, however, there are two strategies that one can imagine to keep the loop working: (i) a large $D_S$ that provides a quick supply of $S$ from neighboring L region holding a higher density of $S$; (ii) a small $D_I$ that the diluting loss of $\rho_{_{A,B}}$ is so slow that the reaction itself is almost off that very low density of $S$ is enough to keep the loop working. This explains why too competitive or too cooperative interaction failed to generate patterns, also the observation that the pattern emergence favors large $D_S$ and small $D_I$. Here the role of baseline infection ratio $R_0$ is similar to $C$, but affecting both initial and secondary infection processes. A small $R_0$ terminates the whole reactions, and a large value on the contrary, results in an imbalance between $S$ and $\rho_{_{AB}}$, therefore a bounded range of $R_0$ is expected to support pattern dynamics.

Obviously, such a dynamical loop is not possible in single contagions, where there is no rich-get-richer mechanism in its dynamics.  This mechanism is seemingly also different from the Turing pattern \cite{turing1952}, where activator and inhibitor species are supposed to present to support the Turing mechanism \cite{kondo2010}. In our systems, there is no well-defined activator or inhibitor. We expect that a systematic classification of our pattern mechanism from the mathematics point of view can be made in future.

\begin{figure*}
\includegraphics[width=0.7\columnwidth]{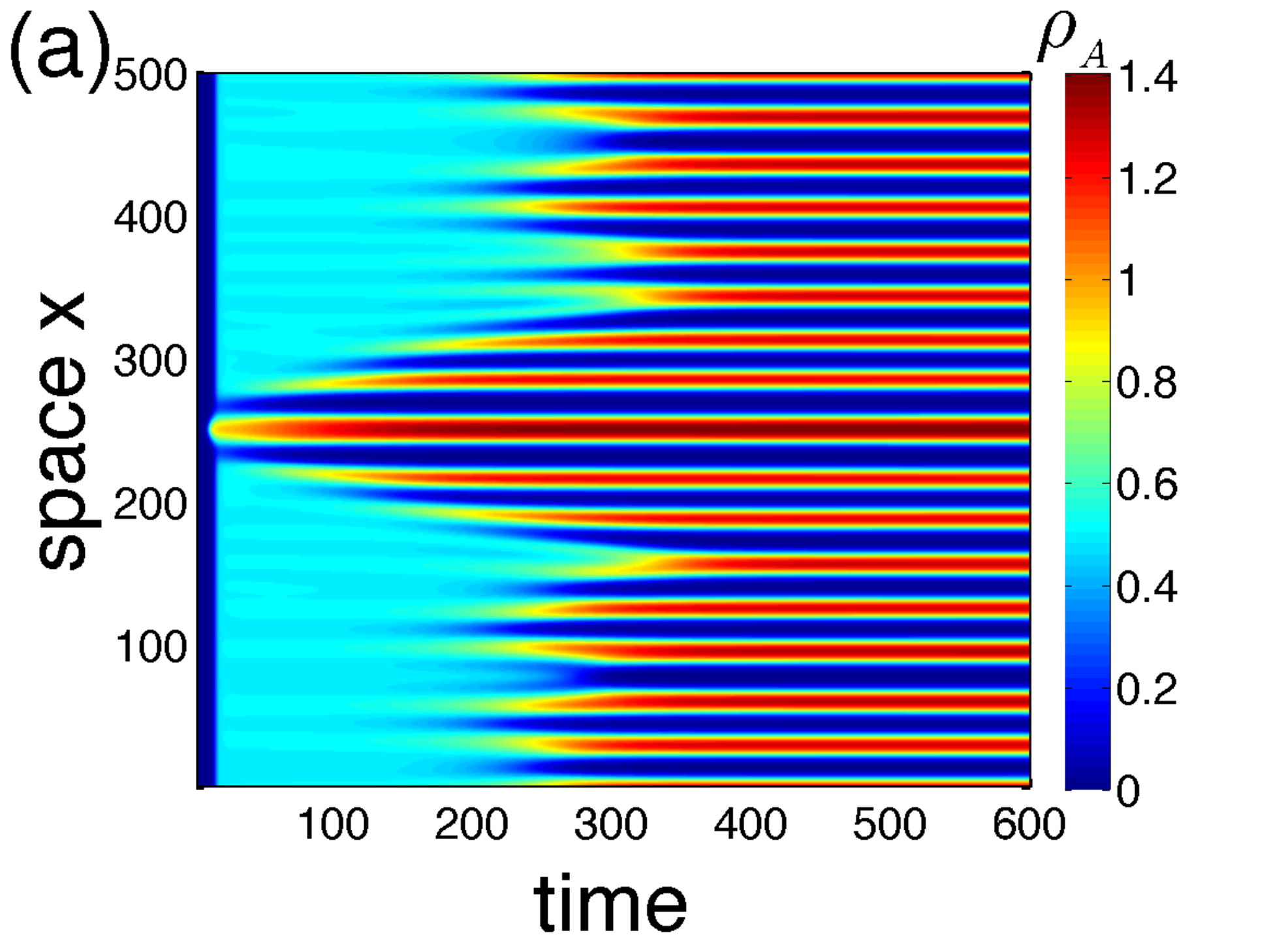}
\includegraphics[width=0.7\columnwidth]{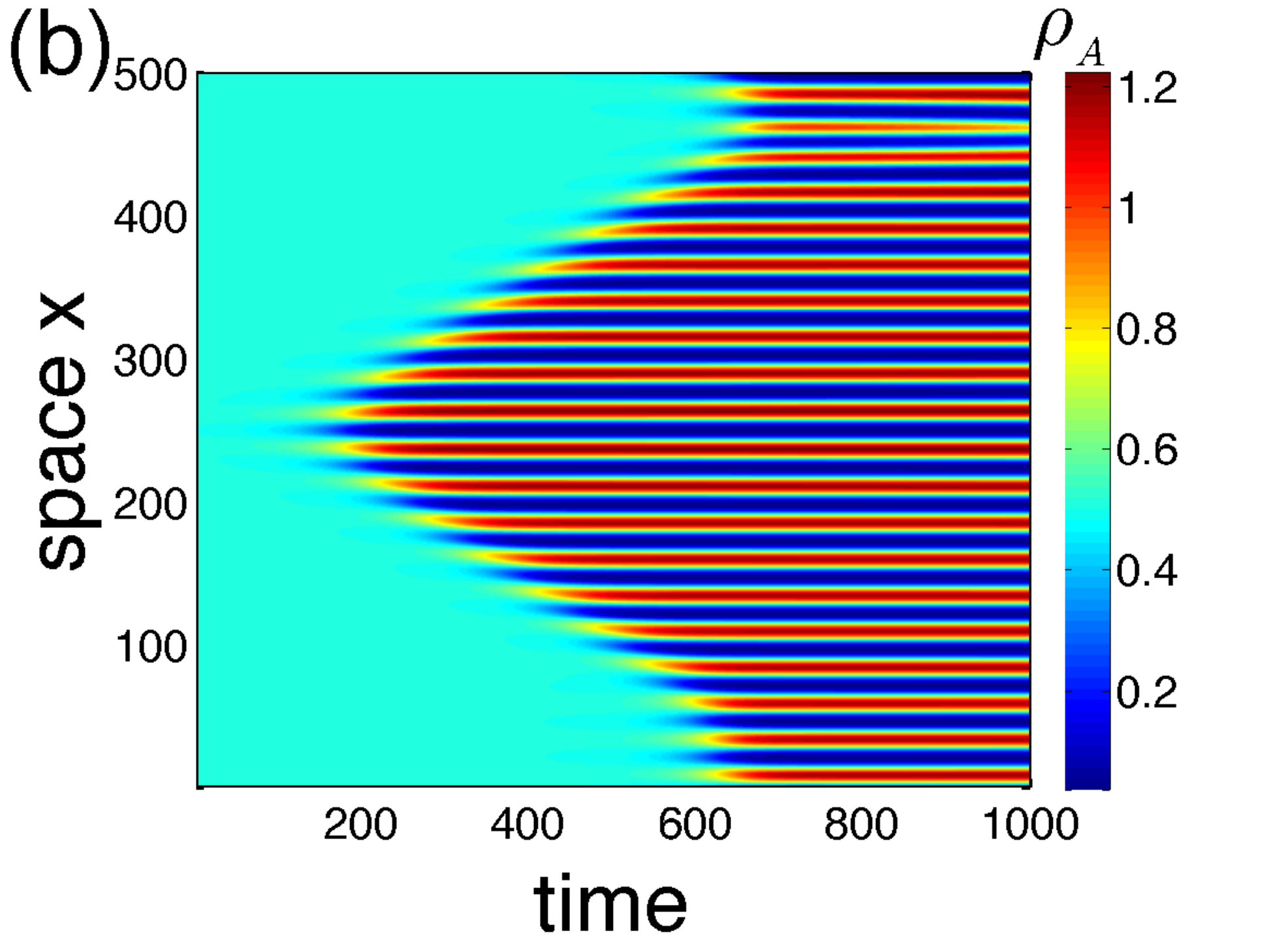}

\includegraphics[width=0.7\columnwidth]{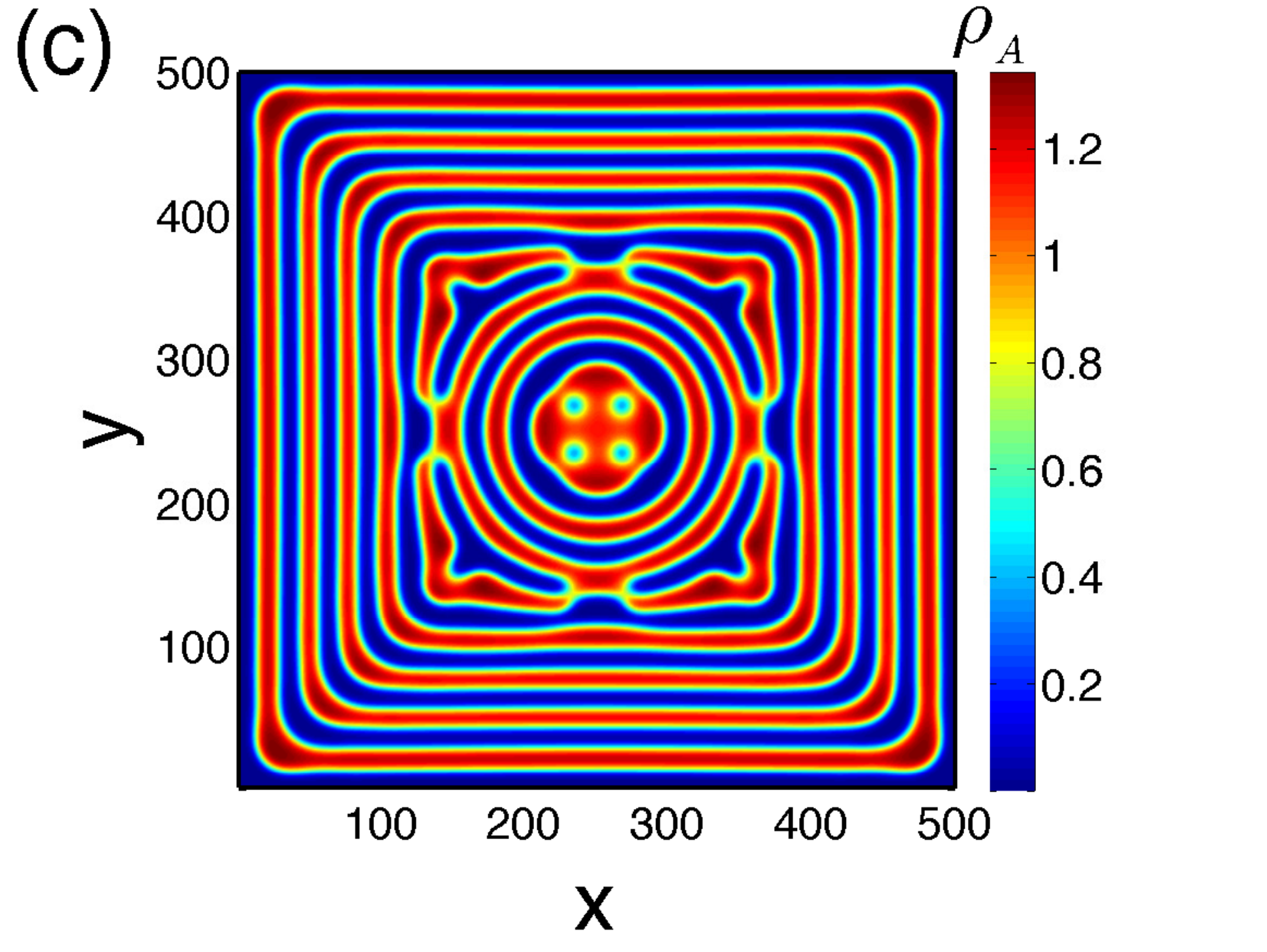}
\includegraphics[width=0.7\columnwidth]{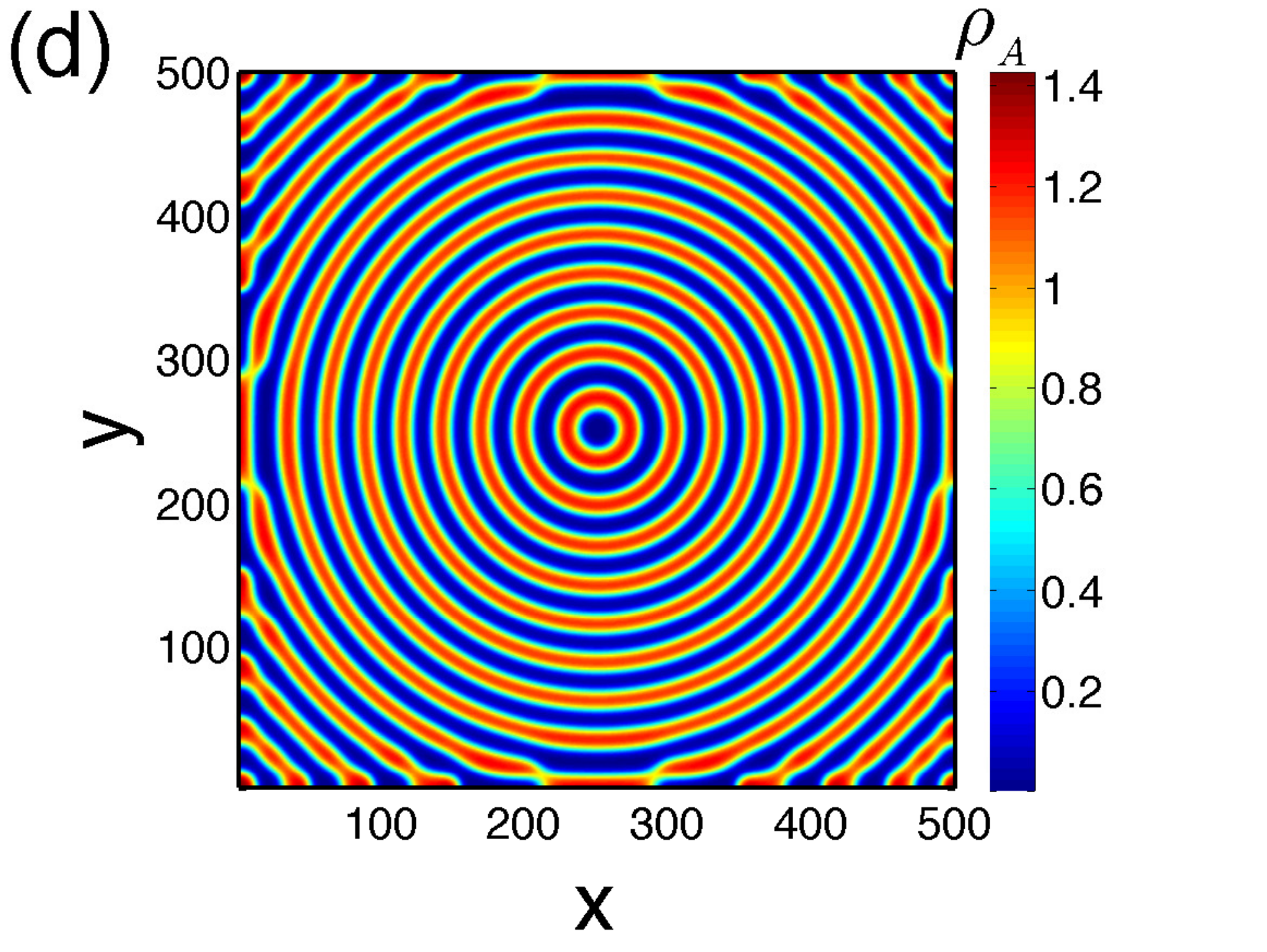}

\caption{\label{fig:9}\textbf{Patterns from a point seed.} (Left column - a, c) Point perturbation in the center
of a complete healthy population; (Right column - b, d) Point perturbation in the center
of a spatially homogeneous population, but in the equilibrium state (here $I_{A}\!=\!I_{B}\!=\!I_{AB}\!=\!1/4$ for the chosen parameters).
The patterns are shown at 500 and 2000 time units, respectively in 
 (c) and (d). Other parameters: $C\!=\!1$, $D_{S}\!=\!10$, $R_{0}\!=\!2$.}
\end{figure*}

\section{Other Aspects} \label{Sec:OA}

\emph{2d Patterns} --- The above observations made in $1d$ space can also be expected in $2d$ domain or even higher, since the linearization analysis does not limited to any specific dimension. As an example, Fig.~\ref{fig:8} shows some typical patterns in $2d$ space starting from random initial conditions, where spots and strips are seen. When the area of strongly infected regions is comparable to the less infected regions, strips emerge. But when the system parameters deviate from this case, a majority/minority composition distributes in the form of spotted patterns. The difference in the two spotted patterns in Fig.~\ref{fig:8} is statistically reversed in highly/weakly infected densities.

\emph{Point seed} --- While most of the simulations above are from random initial conditions, a more realistic scenario would be from a point seed that mimics the importation of few infected individuals into a completely susceptible population. Figure~\ref{fig:9} shows such patterns in $1d$ (Fig.~\ref{fig:9}a) and $2d$ (Fig.~\ref{fig:9}c) domain. While the plots in the left column are for the cases that start from healthy population, the right column start from homogeneous state in equilibrium that are of theoretic interest. We can see that the progression to pattern formation is quite different, and the resulting patterns also share little similarity, \emph{e.g.} the left side is a square pattern while the right is a target pattern in $2d$ domain.

A well-known fact in the study of pattern dynamics is that positive eigenvalues only imply the possibility of pattern in noisy circumstances. The specific shape of patterns is, however, determined by many factors, including both dynamical aspects (such as the reproduction number $R_0$, the interaction $C$, and initial conditions), and the embedded geometry, such as  the shape of the underlining domain and the boundary conditions.

\section{Summary and Discussion}\label{Sec:SD}
In the real world, hundreds or even thousands of different infectious strains simultaneously circulate around and they potentially interact with each other. Here in this study, we try to capture the possible generic contagion scenario by modeling interacting infections in spatial context. Compared to the trivial spatial dynamics of single contagion of classic SIS-type, the presence of more than one agent reveals new contagion complexities. Its emergence does not require any peculiar agent-agent interaction; instead the pattern formation favors mild condition where too strong cooperation or competition is absent. Dynamical mechanism analysis reveals a rich-get-richer phenomenon in the local reaction along with a dynamical loop. This mechanism is rooted in the intrinsic dynamics when two agents get evolved, and too strong agent-agent interaction destroys this loop. Among other observations, one finding of particular interest is pattern hysteresis that the pattern formation is not only determined by the system parameters, but also depends on its evolution direction and history. Since our model is simple enough, only involving two classic SIS infections, we expect some empirical evidences to be found in the future.

The consequence of pattern formation is straightforward that the infection is now spatially segregated; some locations are of high prevalence while their neighborhoods could be much less infected or contagion-free, even though the whole system is in the outbreak phase. The pattern hysteresis in the eradication direction leads to the spatial enhancement in the prevalence, and it implies that the spatial dimension deteriorates the epidemic spreading in the form of contagion pattern. As a direct consequence, a much more effort, if it's not impossible, is required to eradicate the infectious diseases compared to the scenario without space or the single infection case. Here we choose the infectious diseases as our context, where our findings are bad news for healthy departments and for the public, because the minimization or eradication of infections is the primary task. In some other context, however, such as social contagions,  a higher prevalence is usually desired. For example, companies want to sell more products/technologies to their customers; Politicians try to convince more people with their political opinions;  And bloggers want to make their messages a wider readership and more retweets. There, the implications of our study that supports a higher prevalence and stubborn persistence, are actually very good news for them instead.

Our results together with previous related works \cite{chen2013,cai2015,hebert2015,grassberger2016,janssen2016,chen2016} show that the contagion dynamics of two infections is fundamentally different from the classic scenario based on the single infection. These observations of ``more is different" \cite{anderson1972} suggest that realistic contagions could be far more complex than the picture most of previous modeling works captured. 
Besides, our work highlights that the spatial dimension is capable of harboring unexpected amount of complexities in the contagion process, which has largely been underestimated in the past research. In this sense, our work could act as a helpful starting point for a more systematic investigation, and many open questions remain, such as the spread dynamics of more general cases with arbitrary number of agents, how to relate the plain-spaced pattern dynamics to a networked modern world, where heterogeneous transportation systems are often present \cite{brockmann2013}. Other important issues include designing the optimal containment strategies for controlling the prevalence \cite{wang2016}, and the maximization strategies in some other contexts \emph{etc}.




\section*{Acknowledgement}
I would like to thank Dirk Brockmann for his guidance in the early phase of this work, and Olga Baranov and Benjamin Maier in Robert Koch Institute for helpful discussion. Thanks also go to Peter Grassberger for his help in deciphering the possible mechanism behind pattern formation, to Xingang Wang (SNNU) for the discussion at many occasions, and to Ying-Cheng Lai (ASU) and Bernd Sch\"uttler (UGA) for helpful feedback during their visit in Shaanxi Normal University in 2017 summer. This work is supported by the National Natural Science Foundation of China under Grant No. 61703257 and 11747309.

\appendix
\section{Linearisation stability analysis}\label{LSA}
To theoretically analyze the emergence of pattern of Eq. (\ref{eq:rd}), here we follow the standard procedure of linearization stability analysis \cite{murray2003}. Let there be a steady, spatially homogeneous state ($S^{*},I_{A}^{*},I_{B}^{*},I_{AB}^{*}$), which could be an outbreak solution or a contagion-free fixed point of Eq. (\ref{eq:coinfection}), depending on the parameters. The emergence of any non-trivial pattern can be studied by posing perturbations into the system and monitoring the difference regarding to this fixed point, i.e. $(\delta S,\delta I_{A},\delta I_{B},\delta I_{AB})=(S-S^{*},I_{A}-I_{A}^{*},I_{B}-I_{B}^{*},I_{AB}-I_{AB}^{*})$. The evolution of the linearized system can be reformulated by 
\begin{widetext}
\small
\begin{eqnarray}\label{eq:PDE}
\frac{{\partial}}{\partial_{t}}\left(\begin{array}{c}
\delta S\\
\delta I_{A}\\
\delta I_{B}\\
\delta I_{AB}
\end{array}\right) & = & \left(\begin{array}{cccc}
\frac{{\partial f_{s}}}{\partial_{S}}+D_{S}\frac{\partial^{2}}{\partial_{x^{2}}} & \frac{{\partial f_{s}}}{\partial_{I_{A}}} & \frac{{\partial f_{s}}}{\partial_{I_{B}}} & \frac{{\partial f_{s}}}{\partial_{I_{AB}}}\\
\frac{{\partial f_{A}}}{\partial_{S}} & \frac{{\partial f_{A}}}{\partial_{I_{A}}}+D_{I}\frac{\partial^{2}}{\partial_{x^{2}}} & \frac{{\partial f_{A}}}{\partial_{I_{B}}} & \frac{{\partial f_{A}}}{\partial_{I_{AB}}}\\
\frac{{\partial f_{B}}}{\partial_{S}} & \frac{{\partial f_{B}}}{\partial_{I_{A}}} & \frac{{\partial f_{B}}}{\partial_{I_{B}}}+D_{I}\frac{\partial^{2}}{\partial_{x^{2}}} & \frac{{\partial f_{B}}}{\partial_{I_{AB}}}\\
\frac{{\partial f_{AB}}}{\partial_{S}} & \frac{{\partial f_{AB}}}{\partial_{I_{A}}} & \frac{{\partial f_{AB}}}{\partial_{I_{B}}} & \frac{{\partial f_{AB}}}{\partial_{I_{AB}}}+D_{I}\frac{\partial^{2}}{\partial_{x^{2}}}
\end{array}\right)\left(\begin{array}{c}
\delta S\\
\delta I_{A}\\
\delta I_{B}\\
\delta I_{AB}
\end{array}\right).
\end{eqnarray}
\normalsize
Next, we make Fourier transformation, 
\begin{eqnarray}
\delta S^{k} & = & \int\delta S(x,t)e^{-ikx}dx,\\
\delta I_{A}^{k} & = & \int\delta I_{A}(x,t)e^{-ikx}dx,\\
\delta I_{B}^{k} & = & \int\delta I_{B}(x,t)e^{-ikx}dx,\\
\delta I_{AB}^{k} & = & \int\delta I_{AB}(x,t)e^{-ikx}dx,
\end{eqnarray}
where $k$ is the wavenumber. With this operation we reduce the PDEs into ODEs. Inserting the above forms into Eq.  (\ref{eq:PDE}), for a given Fourier mode $k$, we then have
\small
\begin{eqnarray}
\frac{{d}}{dt}\left(\begin{array}{c}
\delta S^{k}\\
\delta I_{A}^{k}\\
\delta I_{B}^{k}\\
\delta I_{AB}^{k}
\end{array}\right) & = & \left(\begin{array}{cccc}
\frac{{\partial f_{s}}}{\partial_{S}}-k^{2}D_{S} & \frac{{\partial f_{s}}}{\partial_{I_{A}}} & \frac{{\partial f_{s}}}{\partial_{I_{B}}} & \frac{{\partial f_{s}}}{\partial_{I_{AB}}}\\
\frac{{\partial f_{A}}}{\partial_{S}} & \frac{{\partial f_{A}}}{\partial_{I_{A}}}-k^{2}D_{I} & \frac{{\partial f_{A}}}{\partial_{I_{B}}} & \frac{{\partial f_{A}}}{\partial_{I_{AB}}}\\
\frac{{\partial f_{B}}}{\partial_{S}} & \frac{{\partial f_{B}}}{\partial_{I_{A}}} & \frac{{\partial f_{B}}}{\partial_{I_{B}}}-k^{2}D_{I} & \frac{{\partial f_{B}}}{\partial_{I_{AB}}}\\
\frac{{\partial f_{AB}}}{\partial_{S}} & \frac{{\partial f_{AB}}}{\partial_{I_{A}}} & \frac{{\partial f_{AB}}}{\partial_{I_{B}}} & \frac{{\partial f_{AB}}}{\partial_{I_{AB}}}-k^{2}D_{I}
\end{array}\right)\left(\begin{array}{c}
\delta S^{k}\\
\delta I_{A}^{k}\\
\delta I_{B}^{k}\\
\delta I_{AB}^{k}
\end{array}\right).
\end{eqnarray}
\normalsize
\end{widetext}

The instability of small perturbations in mode $k$ is then determined by the maximal value of the resulting
eigenvalues $\lambda_{1,2,3,4}^{k}$, where $1,2,3,4$ are from the variable number of the system. Nontrivial
pattern appears if any mode of perturbations is linearly unstable, i.e. $\lambda_{max}=\max_{k}(\lambda_{max}^{k})=\max(\lambda_{1,2,3,4}^{k})>0$. The above analysis can be conveniently extended into a higher spatial dimension by replacing $x$ with $\stackrel{\rightharpoonup}{x}$ without changing the statement at all.

To be specific, when we want to study the impact of any parameter on the pattern dynamics, we will compute the $\lambda_{max}$ as a function of those parameters which are supposed to be already incorporated in A6. Figure ~\ref{fig:3}f shows such an example to examine the role of contagion interaction $C$ in pattern formation. Figure ~\ref{fig:5} is obtained in a similar way, regarding the baseline reproduction number $R_0$. Just in Fig. ~\ref{fig:3}f, with the computed curve of $\lambda_{max} (C)$, we further determine the positive region of  $\lambda_{max}$, where patterns are expected since the homogenous state now becomes unstable. A bit more complicated case is Fig. ~\ref{fig:4}a, where the eigenvalue $\lambda_{max}$ is now a function of both $C$ and $D_S$, and the value of $\lambda_{max}$ is color-coded. To divide the region, a contour line is plotted with value $~ 10^{-5}$ to separate the pattern formation region ($\lambda_{max}\!>\!10^{-5}$) from no pattern region ($\lambda_{max}\!<\!10^{-5}$). The location of the contour line is robust as long as the threshold value is not large, e.g. below $10^{-2}$. In Fig. ~\ref{fig:4}b, we only plot the contour lines for different mobilities of infected $D_I$, with these curves, we can study the impact of $D_I$ on the region available for pattern.

\end{document}